\documentclass[reprint,superscriptaddress,amsmath,amssymb,aps,prl]{revtex4-1}

\usepackage{xcolor}
\usepackage{graphicx}

\usepackage{amsmath}
\usepackage{amsthm}
\usepackage{amssymb}
\usepackage[all]{xy}

\newtheorem{theorem}{Theorem}

\newtheorem{definition}[theorem]{Definition}

\newtheorem{remark}[theorem]{Remark}

\newcommand{\changes}[1]{\textcolor{black}{#1}}
\newcommand{\startchanges}{\color{black}}
\newcommand{\finishchanges}{\color{black}}

\newcommand{\cP}{\mathcal{P}}
\newcommand{\bx}{\mathbf{x}}
\newcommand{\by}{\mathbf{y}}
\newcommand{\bbf}{\mathbf{f}}
\newcommand{\bg}{\mathbf{g}}
\newcommand{\bh}{\mathbf{h}}

\begin{document}

\title{Equitability and explosive synchronisation in multiplex and higher-order networks}

\author{Kirill Kovalenko}
\affiliation{Scuola Superiore Meridionale, School for Advanced Studies, Naples, Italy}

\author{Gonzalo Contreras-Aso}
\affiliation{Grupo Interdisciplinar de Sistemas Complejos (GISC), Departamento de Matem\'aticas, Universidad Carlos III de Madrid, 28911 Legan\'es, Madrid, Spain}

\author{Charo I.~del Genio}
\affiliation{Institute of Smart Agriculture for Safe and Functional Foods and Supplements, Trakia University, Stara Zagora 6000, Bulgaria}
\affiliation{Research Institute of Interdisciplinary Intelligent Science, Ningbo University of Technology, 315104 Ningbo, China}
\affiliation{School of Mathematics, North University of China, 030051 Taiyuan, China}

\author{Stefano Boccaletti}
\affiliation{CNR - Institute of Complex Systems, Via Madonna del Piano 10, I-50019 Sesto Fiorentino, Italy}
\affiliation{Sino-Europe Complexity Science Center, School of Mathematics, North University of China, Shanxi, Taiyuan 030051, China}
\affiliation{Research Institute of Interdisciplinary Intelligent Science, Ningbo University of Technology, Zhejiang, Ningbo 315104, China}

\author{Rubén J.~Sánchez-García}
\affiliation{Mathematical Sciences, University of Southampton, Southampton SO17 1BJ, UK}
\affiliation{Institute for Life Sciences, University of Southampton, Southampton, SO17 1BJ, UK}
\affiliation{The Alan Turing Institute, London, NW1 2DB, UK}

\begin{abstract}
Cluster synchronisation is a key phenomenon observed in networks of coupled dynamical units. Its presence has been linked to symmetry and, more generally, to equability of the underlying pattern of interactions between dynamical units. However, it is not known under which conditions equitability-induced synchronisation is the only cluster synchronisation that can occur on a particular system. Here, we reveal a natural linear independent condition such that equitability becomes necessary, and sufficient, for the existence of cluster synchronised solutions on a very general dynamical system which allows multiplex or higher-order, arbitrarily weighted interactions. Our results explain the ubiquity of explosive synchronisation, as opposed to cluster synchronisation, in multiplex and higher-order networks: equitability imposes additional constraints that must be simultaneously satisfied on the same set of nodes. Our results have significant implications for the design of complex dynamical systems of coupled dynamical units with arbitrary cluster synchronisation patterns and coupling functions. 
\end{abstract}

\maketitle

\section{Introduction}

The study of synchronization in coupled networks of oscillators \cite{boccaletti2006structure} serves as a fundamental pillar to understand the dynamics of interconnected systems, holding profound implications across numerous disciplines \cite{rosenblum2003synchronization, boccaletti2002synchronization, boccaletti2018synchronization}. At its core, the existence of a synchronized state represent how individual elements within a network coordinate their behaviours, fostering coherence or collective dynamics. In recent years, traditional results of global synchronizability in standard networks \cite{pecora1998master} have been extended to both systems with multiple layers of interactions \cite{boccaletti2014structure, della2020symmetries, delgenio2016synchronization, boccaletti2018synchronization, delgenio2022mean} as well as systems with many-body interactions \cite{battiston2020networks, gambuzza2021stability},  constituting a new paradigm that allows researchers to better model the complex processes of real-world systems~\cite{boccaletti2023structure}.


A synchronized state may not involve the whole network. In fact, under specific conditions, one can observe one or more subsets of nodes that synchronize their motion independently of the other groups of nodes and of the rest of the network, giving rise to the phenomenon known as \emph{cluster synchronization}. Notably, even though the nodes within a specific cluster evolve in a coordinated fashion, displaying similar patterns of activity or dynamics, nodes in other clusters may have different synchronization patterns or even remain desynchronized from each other. Understanding how clusters within a network synchronize their activities impacts multiple fields, as it offers insights into the general processes of organization and dynamics in complex systems \cite{pecora2014cluster, schaub2016graph, sorrentino2016complete, siddique2018symmetry, sanchez2020exploiting, jafari2024periodic, sun2025chimera, parastesh2025synchronization}, including brain networks, social networks, power grids, and biological systems. The existence of cluster synchronised solutions has been related to the presence of network symmetries and, more generally, of equitable partitions~\cite{schaub2016graph,o2013observability}, balanced equivalent relations \cite{golubitsky2023dynamics, golubitsky2005patterns, stewart2003symmetry, aguiar2023network} and graph fibrations \cite{stewart2024dynamics, boldi2002fibrations}.

Recently, the study of synchronisation has been considered in generalisations of network models, such as multilayer and hypergraph models \cite{delgenio2016synchronization,gambuzza2021stability}.
Indeed, there is increasing evidence of the importance of higher-order interactions in real-world systems \cite{battiston2020networks, delabays2025hypergraph}, with emerging applications of higher-order dynamics to real-world systems \cite{momani2024directed}. In contrast to network models, explosive synchronisation has been commonly observed in higher-order network dynamics \cite{battiston2021physics, ghorbanchian2021higher}. 

In this article, we formally explain the role of equitability in cluster synchronisation for a very general system of coupled identical dynamical units in multiplex networks (a type of multilayer network with the same nodes on each layer) and hypergraphs (see Fig.~\ref{fig:toyexamples}). We explain why equitability (and thus cluster synchronisation) is much less common in multiplex or higher-order (hypergraph) interactions compared to pairwise (network) interactions. 
Namely, we show that equitability (of the partition of the underlying interaction multiplex, or hypergraph, into clusters) is a necessary condition for the existence of \emph{independent} cluster-synchronised solutions (on those clusters), that is, synchronised solutions for which the coupling functions between clusters are linearly independent over trajectories. Conversely, equitability is also a sufficient condition for synchronisation: we can construct cluster synchronised solutions with respect to any given partition, as long as it is equitable, from the quotient dynamics. The results are independent of the specific form of the coupling functions or of the coupling strength parameters, and depend only of the structural properties (equitability) of the underlying interaction patterns (a network, multiplex, or hypergraph). Although we only concern ourselves with the existence and not the stability of the synchronised solutions (which depends on the specific form of the coupling functions and, for fixed coupling functions, on the values of the coupling strength parameters), our results show how equitability gives strong restrictions on the existence of synchronised solutions, and, in particular, how the restrictions are much harder to achieve for higher-order dynamical interactions.

\section{Results}
The results section is organised as follows. After describing the dynamical systems, we first study the multiplex (over the next few sections), then the hypergraph case when the coupling functions are non-invasive, which includes the diffusive case and relates to \emph{external} equitability. Then we explain how to generalise our results to general coupling functions, which relates to \emph{general} equitability, and why both cases must be treated separately. We finish with a discussion on explosive synchronisation, and on finding equitable partitions. 

\subsection{Dynamical system and cluster synchronisation}

\begin{figure*}[t]
    \centering
    \includegraphics[width=0.9\linewidth]{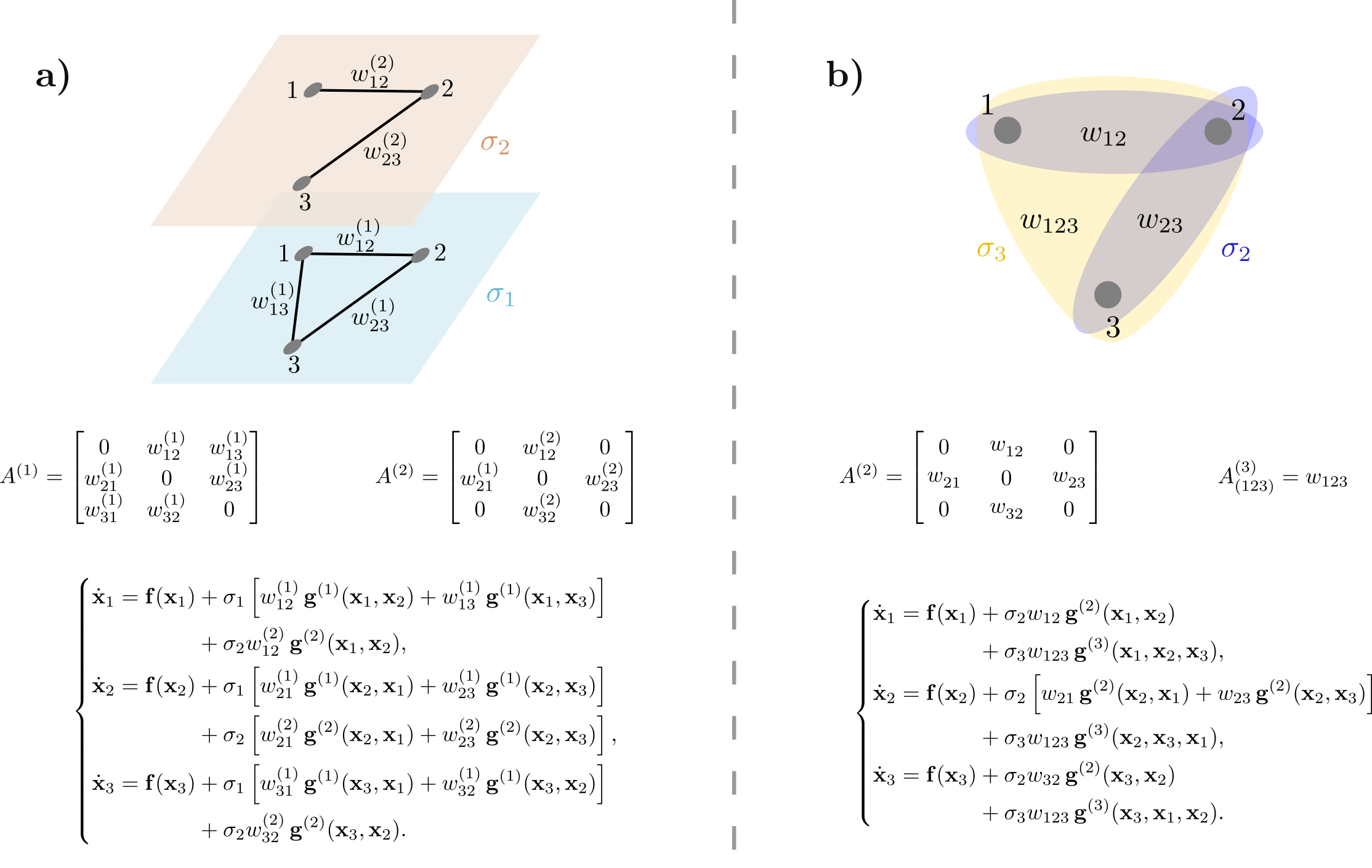}
    \caption{\textbf{Multiplex and hypergraph dynamical systems.} Toy example of $\mathbf{a)}$ a biplex (multilayer network with two layers with three identical nodes), and $\mathbf{b)}$ a hypergraph with pairwise and triadic interactions among three nodes, along with their respective adjacency matrices/tensors and dynamical equations. The shorthand $(123)$ indicates any permutation of the three indices. To illustrate equitability, consider the partition into clusters $\{1, 3\}$ and $\{2\}$: it is externally equitable for the multiplex $\mathbf{(a)}$ if $w_{12}^{(1)}=w_{23}^{(1)}$ (layer 1) and $w_{12}^{(2)}=w_{23}^{(2)}$ (layer 2); and for the hypergraph $\mathbf{(b)}$ if $w_{12}=w_{23}$.}
    \label{fig:toyexamples}
\end{figure*}

The behaviour of a networked system of $N$ identical $D$-dimensional units is described by a system of dynamical equations of the form (see SI for full details)
\begin{equation}\label{eq:dynamical-system-general}
    \dot{\bx}_i = \bbf(\bx_i) + \mathbf{h}_i(\bx_{1},\dots, \bx_N) \qquad i=1,\ldots,N
\end{equation}
where $\bx_i(t)$ is a $D$-dimensional vector describing the state of the $i$th unit at time $t\in T \subseteq \mathbb{R}$, $\bbf$ is a smooth map that encodes the internal dynamics of the units, and $\bh_i(\bx_{1},\dots, \bx_N)$ is a coupling term, representing the total dynamical input of all other nodes to node $i$. In a multiplex network with $M$ layers (and the same $N$ nodes on each layer), the coupling is of the general form 
\begin{equation}\label{eq:coupling-multilayer}
 \bh_i^{\mathrm{multi}} = \sum_{m=1}^M\sigma_m\sum_{j=1}^N A^{(m)}_{i,j}\mathbf g^{(m)}(\bx_i, \bx_j)\:,
\end{equation}
whereas in a higher-order network (hypergraph) with highest multi-body interactions order~$M$, it is \cite{gambuzza2021stability}
\begin{multline} \label{eq:coupling-hypergraph}
 \bh_{i}^{\mathrm{hyper}} = \sum_{m=1}^M\sigma_m\\ \sum_{j_1,\ldots,j_m=1}^N A^{(m)}_{i,j_1,\dotsc,j_m}\mathbf g^{(m)}(\bx_{i}, \bx_{j_1}, \dotsc, \bx_{j_m})\:.
\end{multline}
(See Fig.~\ref{fig:toyexamples} for some toy examples.)
In the equations above, $\sigma_m > 0$ is the coupling strength for layer~$m$ or order~$m$, $A^{(m)}$ is the adjacency matrix for layer $m$ (in the multiplex case), or the hypergraph adjacency tensor of order (number of elements in a relation, in the hypergraph case) $m+1$, and the coupling functions~$\mathbf g$ are assumed to be synchronization non-invasive, that is, $\mathbf g^{(m)}(\bx, \bx, \dotsc, \bx)=0$ for all $m$ and any $\bx$. Non-invasiveness \cite{jungling2015synchronization, gambuzza2021stability} guarantees that the coupling does not alter the intrinsic dynamics of the dynamical units when they are synchronised, and it is a mild generalisation of the class of diffusive systems, that is, those in which coupling functions depend (possibly linearly) on differences between states. Our results, however, can be adapted to generic coupling functions beyond non-invasive ones; this needs to be treated as a separate case, explained in the section `General coupling' after discussing the non-invasive multiplex, and hypergraph, cases.
For $M=1$, both the multiplex and the hypergraph case reduce to the network (or monolayer) case, 
\begin{equation}\label{eq:dynamical-system-network}
    \dot{\bx}_i = \bbf(\bx_i) + \sigma \sum_{j=1}^N a_{ij} \, \mathbf{g}(\bx_i, \bx_j),
\end{equation}
where $a_{ij}$ are the entries of the network adjacency matrix $A=(a_{ij})$. An important example in this case is the graph Laplacian coupling, $\bg(\bx_i, \bx_j) = \bx_j - \bx_i$. 



A solution $\vec\bx = (\bx_1,\ldots,\bx_N)$ to the dynamical system Eq.~\eqref{eq:dynamical-system-general} is \emph{globally synchronised} if $\bx_1 = \ldots = \bx_N$ (identical synchronisation). Since $\bg$ is non-invasive, this is equivalent to the synchronised solution $\bx_s = \bx_i$ being a solution of $\dot{\bx} = \bbf(\bx)$. Many systems, however, exhibit \emph{partial synchronisation}, that is, identical synchronisation in a subset, or several subsets, of nodes or dynamical units \cite{pecora2014cluster, boccaletti2018synchronization, boccaletti2002synchronization}. Formally, given a \emph{partition} $\mathcal{P}$ of the node set $V$ (a collection of pairwise disjoint subsets of $V$, called \emph{clusters}, whose union is $V$), we say that a solution $\vec\bx$ is \emph{$\mathcal{P}$-synchronised} if 
\begin{equation}\label{eq:cluster-synchronisation}
    \bx_i = \bx_j \text{ for all } i, j \in C,
\end{equation}
for each $C \in \cP$. 
This includes any form of synchronisation (that is, involving any subsets of nodes), as we can put single nodes on (trivially) synchronised clusters on their own --- we call such cluster a \emph{singleton}. We note that, if the partition has $k$ clusters, the case $k=1$ corresponds to global synchronisation and the case $k=n$ to no synchronisation.

In this article, we are concerned with very general necessary and sufficient conditions for the \emph{existence} of cluster synchronised solutions, based on the equitability properties of the underlying pattern of interactions. We will not make any assumptions or claims on the \emph{stability} of the solutions, which typically depends on the actual form of the dynamical system, such as the coupling functions $\bg^{(m)}$, or, for given coupling functions, the values of the coupling strength parameters $\sigma_m$ (cf.~Fig.~\ref{fig:multilayer-condition-fulfilled-and-not}).

For simplicity, we focus on the multiplex case first, and discuss the hypergraph case later, in the Section called `Results for hypergraph dynamics'.

\subsection{Dynamical and external equitability}
We want to relate cluster (or partial) synchronisation to equitability, that is, the fact that every node receives the same input from nodes in other clusters. We first introduce a notion of dynamical equitability, then relate this to a more standard notion of external equitability \cite{schaub2016graph, sorrentino2016complete, o2013observability}. 

First, we introduce some notation. Given a subset of nodes $C \subseteq V$ and a solution $\vec\bx = (\bx_1,\ldots,\bx_N)$ of the dynamical system, we write 
\begin{align}
    \bh_i^{C} &= \sum_{m=1}^M \sigma_m \sum_{j \in C} A_{i,j}^{(m)} \bg^{(m)}(\bx_i,\bx_j),
\end{align}
the total dynamical input to node $i$ from nodes in cluster $C \subseteq V$. We can see $\bh_i^{C}$ as a function of $\bx_1,\ldots,\bx_N$, or as a function of $t$ as $\bx_j = \bx_j(t)$ for all $j$. Note that $\bh_i^{C}$ is a non-invasive function of $\bx_1,\ldots,\bx_N$, since all the coupling functions $\bg^{(m)}$ are non-invasive, and additive, in the sense that $\bh_i^{A \cup B} = \bh_i^{A} + \bh_i^{B}$ whenever $A$ and $B$ are disjoint clusters. 

We can now rewrite Eq.~\eqref{eq:dynamical-system-general} in the multiplex case in terms of the internal and external dynamical inputs with respect to a partition $\cP$, 
\begin{align}
    \dot\bx_i = f(\bx_i) + \bh^\text{int}_i + \bh^\text{ext}_i,
\end{align}
where $i \in C$, $C \in \cP$, $\bh^\text{int}_i = \bh_i^C$, and $\bh^\text{ext}_i = \bh_i^{V\setminus C}$. In particular, if $\vec\bx$ is a $\cP$-synchronised solution, the internal dynamical input to each node is zero ($\bh^\text{int}_i(t) = 0$ all $t \in T$, all $i$) since the coupling functions are non-invasive, and the external contribution to two nodes in the same cluster is equal, that is, 
\begin{align}\label{eq:weak-external-dynamical-equitability}
    \bh^\text{ext}_i = \bh^\text{ext}_j
\end{align}
whenever $i, j \in C$, $C \in \cP$. This motivates the following definition of dynamical equitability. We call a solution $\vec\bx$ \emph{dynamically $\cP$-equitable} if Eq.~\eqref{eq:weak-external-dynamical-equitability} is satisfied for all $i, j \in C$, for all $C \in \cP$. Thus we have showed that dynamical equitability is a necessary condition for cluster synchronisation. 

We would like to relate this notion to a more standard notion of external equitability, that is, equitability with respect to the interaction pattern of the nodes in the underlying multiplex network. If $C \subseteq V$, we write 
\begin{equation}
    h_i^{C,m} = \sum_{k \in C} A^{(m)}_{ik} 
\end{equation}
for the structural input to node $i$ from all nodes in $C$ on layer $m$. This is a constant that depends only on the partition $\mathcal{P}$ and the underlying interaction multiplex network, but not on the coupling functions $\bg^{(m)}$ nor on a solution $\bx$. We say that $\mathcal{P}$ is \emph{externally equitable} if, for all $C, C' \in \mathcal{P}$, $C \neq C'$, and $1 \le m \le M$, 
\begin{equation}\label{eq:structural-equitability}
    h_i^{C',m} = h_j^{C',m} 
\end{equation}
for all $i, j \in C$. This means that the total, or aggregated, input to a node in $C$ from all nodes in $C'$ is the same (see Fig.~\ref{fig:toyexamples}).
The condition $C \neq C'$ results in `external' equitability, and is included since the internal connectivity in each cluster does not play a role in cluster synchronisation, as shown above, due to the non-invasiveness of the coupling function (which implies $\bh_i^\text{int}=\mathbf{0}$ for all $i$). 
Note that condition \eqref{eq:structural-equitability} is always satisfied when $\cP$ is the partition into one cluster (the condition becomes empty and thus true by convention) or $N$ clusters.

It is easy to show that external equitability implies dynamical equitability for any cluster synchronised solution (see SI). We will show that, under a natural cluster linear independence condition (next section), dynamical and external equitability are in fact equivalent for cluster-synchronised solutions, and relate them more generally to cluster synchronisation (in the two sections after that). 



In the SI, we study several definitions of dynamical and external equitability and their relationships in more detail, and justify our choices above. 

\subsection{Independent cluster synchronisation}

We call a cluster-synchronised solution \emph{independent} if the coupling functions (across clusters, and layers) with all other clusters are linearly independent over trajectories. Formally, given $\cP$ a partition and $\vec\bx$ a $\cP$-synchronised solution, we call the solution $\vec\bx$ $\cP$-\emph{independent} if, for each $C \in \cP$, the $M \cdot (|\cP|-1)$ functions 
\begin{equation}\label{eq:cluster-independence}
    \bg^{(m)}(\bx_C,\bx_{C'}), \ m=1,\ldots,M, \  
    C' \in \cP, \ C' \neq C,
\end{equation}
are linearly independent over the time domain $T$, where we write $\bx_C(t)$ for the synchronised solution on $C$ (i.e.~$\bx_C(t)=\bx_i(t)$ for all $i \in C$), and similarly for $\bx_{C'}$ (see the SI for full details). We call this \emph{(linearly) independent cluster synchronisation}: cluster synchronisation (Eq.~\eqref{eq:cluster-synchronisation}) with linearly independent coupling functions over trajectories (Eq.~\eqref{eq:cluster-independence}). \changes{We note that Eq.~\eqref{eq:cluster-independence} is always satisfied if $\cP$ is the partition into one cluster (as empty conditions are true by convention) or $N$ clusters (singleton partition), corresponding to global respectively no synchronisation. }

In some sense, independent cluster synchronisation is `generic': clusters are independent unless they are synchronised ($\bx_C = \bx_{C'}$, which we can resolve by merging synchronised clusters together), or there is a  linear relation between them. Namely, the following are the different ways in which a solution may fail to be independent with respect to a partition: 
\begin{itemize}
    \item[(1)] there are \emph{synchronised clusters}, that is, $\bx_C = \bx_{C'}$ for some $C \neq C'$, which immediately implies $\bg^{(m)}(x_C,x_{C'}) = \mathbf{0}$ identically 0;
    
    \item[(2)] the \emph{coupling} between two clusters is \emph{zero} despite not being (identically) synchronised, that is, $\bg^{(m)}(x_C,x_{C'}) = \mathbf{0}$ even though $\bx_C \neq \bx_{C'}$; 
    
    (As an example, consider a sinusoidal Kuramoto coupling function 
    \begin{equation}
        \bg^{(m)}(\bx_i,\bx_j) = \sin(\bx_i-\bx_j)
    \end{equation}
    and a solution with $\bx_C(t) = \bx_{C'}(t) + k \pi$ for an integer $k$.)
    
    \item[(3)] there are \emph{linearly dependent layers}, for example $\bg^{(1)} = \bg^{(2)} + \bg^{(3)}$ on some trajectory $(\bx_C(t), \bx_{C'}(t))$, even if $\bg^{(1)} \neq \bg^{(2)} + \bg^{(3)}$ in general;
    
    (For an example, consider the coupling functions 
    \begin{equation}
        \begin{cases}
            \bg^{(1)}(x_1,x_2)&=4x_1-1,\\
            \bg^{(2)}(x_1,x_2)&=x_1+x_2,\\
            \bg^{(3)}(x_1,x_2)&=x_1+x_2+1,
        \end{cases}
    \end{equation} 
    with the trajectories given by 
    \begin{equation}
        \begin{cases}
            x_C(t)=t,\\
            x_{C'}(t)=t-1,
        \end{cases}
    \end{equation}
    then $\bg^{(1)} \neq \bg^{(2)} + \bg^{(3)}$ in general but $\bg^{(1)}(t) = \bg^{(2)}(t) + \bg^{(3)}(t)$ over the trajectory $(x_{C}(t), x_{C'}(t))$.)
    
    \item[(4)] there are \emph{linearly dependent clusters within a layer}, for example $\bg^{(m)}(\bx_C,\bx_{C_1}) = \bg^{(m)}(\bx_C,\bx_{C_2}) + \bg^{(m)}(\bx_C,\bx_{C_3})$;

    (As an example, consider the coupling function $\bg^{(m)}(x_1,x_2)=x_1+x_2$ and cluster solutions $\bx_C(t)=t$, $\bx_{C_1}(t)=6t$, $\bx_{C_2}(t)=2t$, and $\bx_{C_3}(t)=3t$.)
\end{itemize}
and any combination of the above. \changes{Later, we will see that these are the only cases in which cluster synchronisation (and thus dynamical equitability) can occur without requiring external equitability of the underlying partition.} 

All but case (1), which simply means that synchronised nodes need to be in the same cluster, depend on the specific form of the $M$ coupling functions $\bg^{(m)}$, and in particular on their algebraic properties, and thus we cannot say anything more in general. From now on, therefore, we will focus on independent cluster synchronisation.

\subsection{Independent cluster synchronisation implies external equitability}
Our first result is that, under independent synchronisation, dynamical and external equitability (see previous section) are equivalent. That is, if $\vec\bx$ is a $\cP$-synchronised (Eq.~\eqref{eq:cluster-synchronisation}), $\cP$-independent (Eq.~\eqref{eq:cluster-independence}) solution, then the following are equivalent:
\begin{itemize}
    \item[(i)] $\vec\bx$ is dynamically $\cP$-equitable (Eq.~\eqref{eq:weak-external-dynamical-equitability});
    \item[(ii)] $\cP$ is externally equitable (Eq.~\eqref{eq:structural-equitability}). 
\end{itemize}
(See the SI for a full derivation.) Since we have shown that dynamically equitability is a necessary condition for (linearly independent) cluster synchronisation, so is external equitability. This restricts the groups of nodes in a network that can support cluster synchronisation to those forming equitable partitions simultaneously on each layer. (In the next section, we will see that external equitability is also a sufficient condition for cluster synchronisation.) 

\changes{We could have considered weaker, less restrictive, notions of external equitability, for instance }equitability with respect to all layers instead of per layer, or with respect to all external nodes instead of per cluster, but we can show that they are \emph{not} necessary for cluster synchronisation (see SI for a full comparison on difference notions of equitability). Therefore, in the context of independent cluster synchronisation, a partition must be equitable \emph{on each layer simultaneously}. 
This explains why cluster synchronisation is much harder to achieve in multi-layer networks: the equitability condition must be simultaneously satisfied on each layer, for the same nodes. \changes{Indeed, even if many networks have equitable partitions with high probability somewhere on the network (for example, orbit partitions in real-world networks \cite{macarthur2008symmetry, sanchez2020exploiting}), the probability of equitable clusters to occur on the same nodes in two, or more, layers decays very rapidly with the number of layers (see the SI for an explicit calculation).} 

\begin{figure*}
    \centering
    \includegraphics[width=0.9\linewidth]{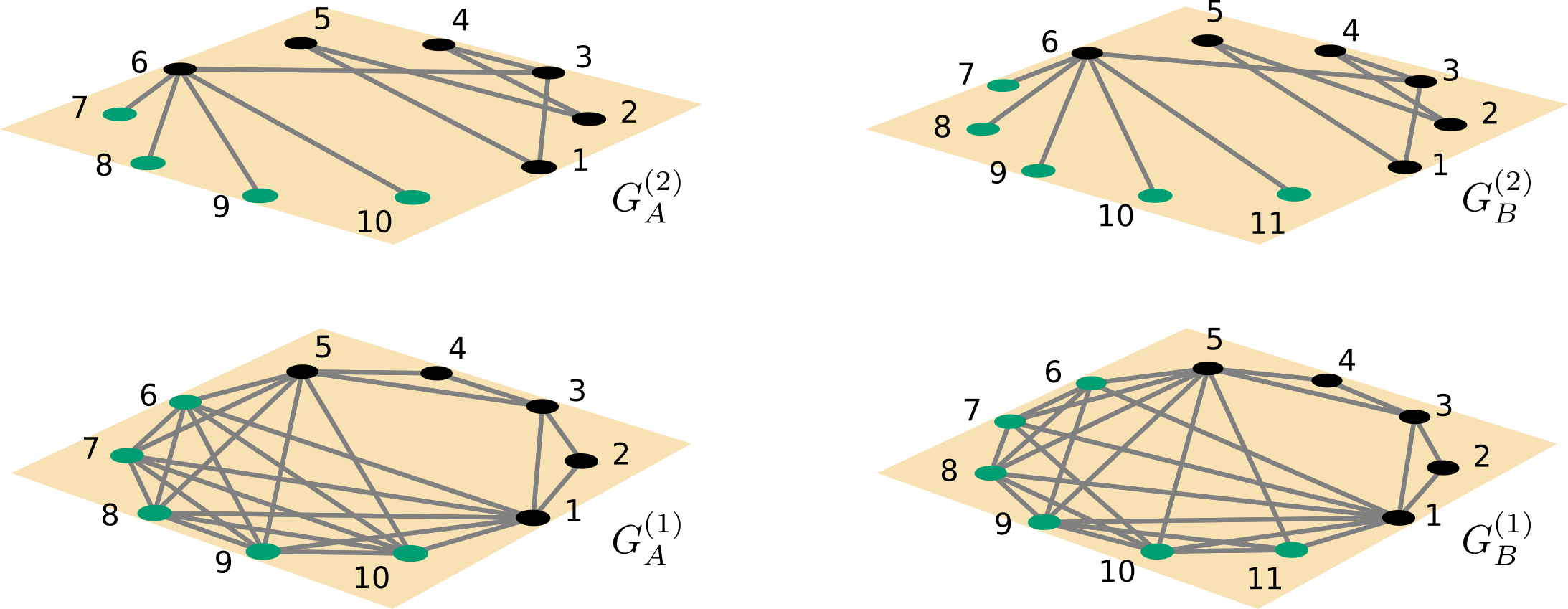}
    \caption{\textbf{Equitability on multiplex networks.} Two biplex networks $G_A=(G_A^{(1)}, G_A^{(2)})$ on the left and $G_B=(G_B^{(1)}, G_B^{(2)})$ on the right that exhibit very different dynamical behaviour: cluster synchronization on $G_A$, and explosive synchronization on $G_B$ (see Fig.~\ref{fig:multilayer-condition-fulfilled-and-not} for the accompanying numerical results). This can be easily explained in terms of equitability: the green nodes form an equitable cluster in their respective layers, but only the common nodes $\{7,8,9,10\}$ are equitable overall on $G_A$, with no similar equitable partition on $G_B$.}
    \label{fig:multilayer-condition-fulfilled-and-not-graphs}
\end{figure*}

Let us illustrate the simultaneous external equitability condition with the two multilayer networks shown in Fig.~\ref{fig:multilayer-condition-fulfilled-and-not-graphs}. These are biplex networks, that is, multiplex networks (same nodes on all layers) with two layers. The first biplex network $G_A=(G_A^{(1)}, G_A^{(2)})$ has a partition which is externally equitable in both layers simultaneously: the one with nodes $C=\{7, 8, 9, 10\}$ as one cluster and all other nodes as singletons. Later, we show that cluster synchronisation does occur on $C$, on a numerical example (Fig. \ref{fig:multilayer-condition-fulfilled-and-not}). On the other hand, the second biplex network $G_B=(G_B^{(1)}, G_B^{(2)})$ violates the simultaneous equitable condition: even though there are almost identical  external equitable partitions for every layer individually, $\{6, 7, 8, 9, 10, 11\}$ on layer 1 and $\{7, 8, 9, 10, 11\}$ on layer 2, there are no partitions which are equitable on both layers simultaneously (node 6 is essential to achieve equitability on layer 1, but prevents equitability on layer 2). Again, we show later on a numerical example (Fig.~\ref{fig:multilayer-condition-fulfilled-and-not}) that only explosive synchronisation is observed in this case. 

Altogether, we have shown that cluster synchronisation can only occur in one of these cases:

\begin{itemize}
    \item there is a linear dependence relation between the coupling functions on trajectories (as illustrated in the previous section); or, if not, 
    \item the partition of the underlying interaction network must be layer-by-layer externally equitable (Eq.~\eqref{eq:structural-equitability}). 
\end{itemize}
This summarises the key insight of our article: unless there are linear dependencies between coupling functions (on trajectories), cluster synchronisation can only occur if the underlying partition is externally equitable layer by layer. \changes{(For a comparison to other possible notions of equitability, see the SI.)} 

In the next section, we show the converse statement, that is, that external equitability (layer by layer) is not only necessary, but sufficient to guarantee the existence of a synchronised solution, with the help of the quotient network formalism.


Before moving on, we briefly remark on external equitability and its relation to symmetry. Network symmetries (formally, permutation of the nodes preserving (weighted) edges) are an important source of equitable partitions, namely partitions into orbits of structurally equivalent nodes \cite{macarthur2008symmetry, sanchez2020exploiting}. For multiplex networks, and also for hypergraphs, we can extend the definition to permutations of the nodes preserving edges on each layer, respectively hyperedges of the same order. As an example, the equitable partition in the biplex $G_A$ in Fig.~\ref{fig:multilayer-condition-fulfilled-and-not-graphs} is indeed a symmetry partition into orbits. 

\subsection{Quotient dynamics}
\changes{Quotients with respect to equitable partitions have been studied in the context of cluster synchronisation and dynamics e.g.~\cite{golubitsky2005patterns, sorrentino2019symmetries, aguiar2023network}. Our treatment of quotient and quotient dynamics here is similar or equivalent to that found elsewhere.} 

If $A$ is an $N \times N$ real matrix, say the adjacency matrix of a weighted, directed network with $N$ nodes, and $\cP = \{C_1, \ldots, C_K\}$ is a partition of the node set $\{1,\ldots, N\}$ into $K$ clusters, the \emph{quotient} of $A$ with respect to $\cP$ is the $K \times K$ matrix $B$ with $(k,l)$-entry
\begin{equation}
    B_{k,l} = \frac{1}{|C_k|} \sum_{i \in C_k, j \in C_l} A_{i,j},
\end{equation}
the average connectivity from all nodes in $C_l$ to a node in $C_k$. 

Given the dynamical system (Eqs.~\eqref{eq:dynamical-system-general} and \eqref{eq:coupling-multilayer}) and a partition $\cP$ of the dynamical units (equivalently, of the set $\{1,\ldots,N\}$) into $K$ clusters, we define its \emph{quotient dynamical system} with respect to $\cP$ as the same dynamical system (same internal dynamics $\bbf$, coupling functions $\bg^{(m)}$, and coupling strengths $\sigma_m$) on the quotient multiplex network, that is, the multiplex network with quotient adjacency matrices $B^{(m)}$, $m = 1,\ldots, M$. Explicitly, we have dynamical units $\by_1,\ldots, \by_K$ and $K$ differential equations 
\begin{equation}\label{eq:quotient-dynamics}
    \dot{\by}_i = \bbf(\by_i) + \bh_i(\by_1,\ldots,\by_K),
\end{equation}
where 
\begin{equation}
    \bh_i(\by_1,\ldots, \by_K) = \sum_{m=1}^M \sigma_m \sum_{j=1}^K B^{(m)}_{i,j}\, \bg^{(m)}(\by_i,\by_j).
\end{equation}
We refer to this dynamical system as the \emph{quotient} dynamics, and to the original one as the \emph{parent} dynamics. 

In the SI, we show a correspondence between $\cP$-synchronised $\cP$-equitable solutions $\vec\bx$ of the parent dynamical system, and (global) solutions $\vec\by$ of the quotient dynamical system, by setting $\by_k = \bx_i$ for any $i \in C_k$. Formally, $\bx=(\bx_1,\ldots,\bx_N)$ is a $\cP$-synchronised, $\cP$-equitable solution of Eq.~\eqref{eq:dynamical-system-general} if and only if $\vec\by=(\by_1,\ldots,\by_M)$ is a solution of Eq.~\eqref{eq:quotient-dynamics}, where $\by_k = \bx_i$ for any $i \in C_k$. 

This means that we can construct equitable, cluster synchronised solutions on an arbitrary multiplex network with respect to an arbitrary partition, simply by `lifting' (repeating on each node in a cluster) a solution of the quotient multiplex network. 
That is, in this sense, equitability implies cluster synchronisation (the existence of cluster synchronised solutions on the clusters of the chosen equitable partition).
Again, we are not making any claims on the \emph{stability} of the synchronised solutions: even if the quotient solution is stable, the parent solution will be stable under uniform perturbations of all the nodes in the cluster, but not necessarily under arbitrary, non-uniform perturbations. 

As an example, the biplex network $G_A$ in Fig.~\ref{fig:multilayer-condition-fulfilled-and-not-graphs} was constructed in this way: starting with an arbitrary (quotient) network on 7 nodes, substitute node 7 by an arbitrary graph (on four nodes 7, 8, 9 and 10 in this case) and connect the new nodes to the other nodes in an equitable way (each node 1 to 6 connects to either all, or none, of the nodes 7 to 10 in the cluster) to create the parent graph. (Of course, other, more complex, equitable connectivity patterns are possible.) 






\subsection{Numerical Results} \label{sec:numerical}

We illustrate the relation between equitability and cluster synchronisation with numerical simulations of coupled Lorenz oscillators on the two biplex network shown in Fig.~\ref{fig:multilayer-condition-fulfilled-and-not-graphs} as we vary the coupling strength parameters $\sigma_1$ and $\sigma_2$. \changes{The quantity of interest we track here is the synchronization error, 
\begin{equation}\label{eq:syncerror}
    S_k = \left\langle \sqrt{\frac{1}{N_k} \sum_{i\in C_k} \left|\bx_i - \overline{\bx}_{k}\right|^2} \right\rangle_{T},
\end{equation}
where $C_k$ is the set of nodes contained in cluster $k$ (or all nodes in the case of the total error), $\overline{\bx}_{k}$ is the ensemble average of $\vec\bx= (\bx_1,\ldots,\bx_N)$ within $C_k$, $\overline{\bx}_k = \frac{1}{N_k}\sum_{i\in C_k} \bx_i$, where $N_k = |C_k|$ the number of nodes in $C_k$, and $\langle \cdot \rangle_{T}$ denotes the temporal average over a late time window $T$.} Full details of the simulations, including the dynamical functions involved, which belong to Class II \cite{boccaletti2006structure} on each layer (global synchronisation guaranteed for high enough values of $\sigma_i$), can be found in the SI.  

\begin{figure*} 
    \centering
    \includegraphics[width=\linewidth]{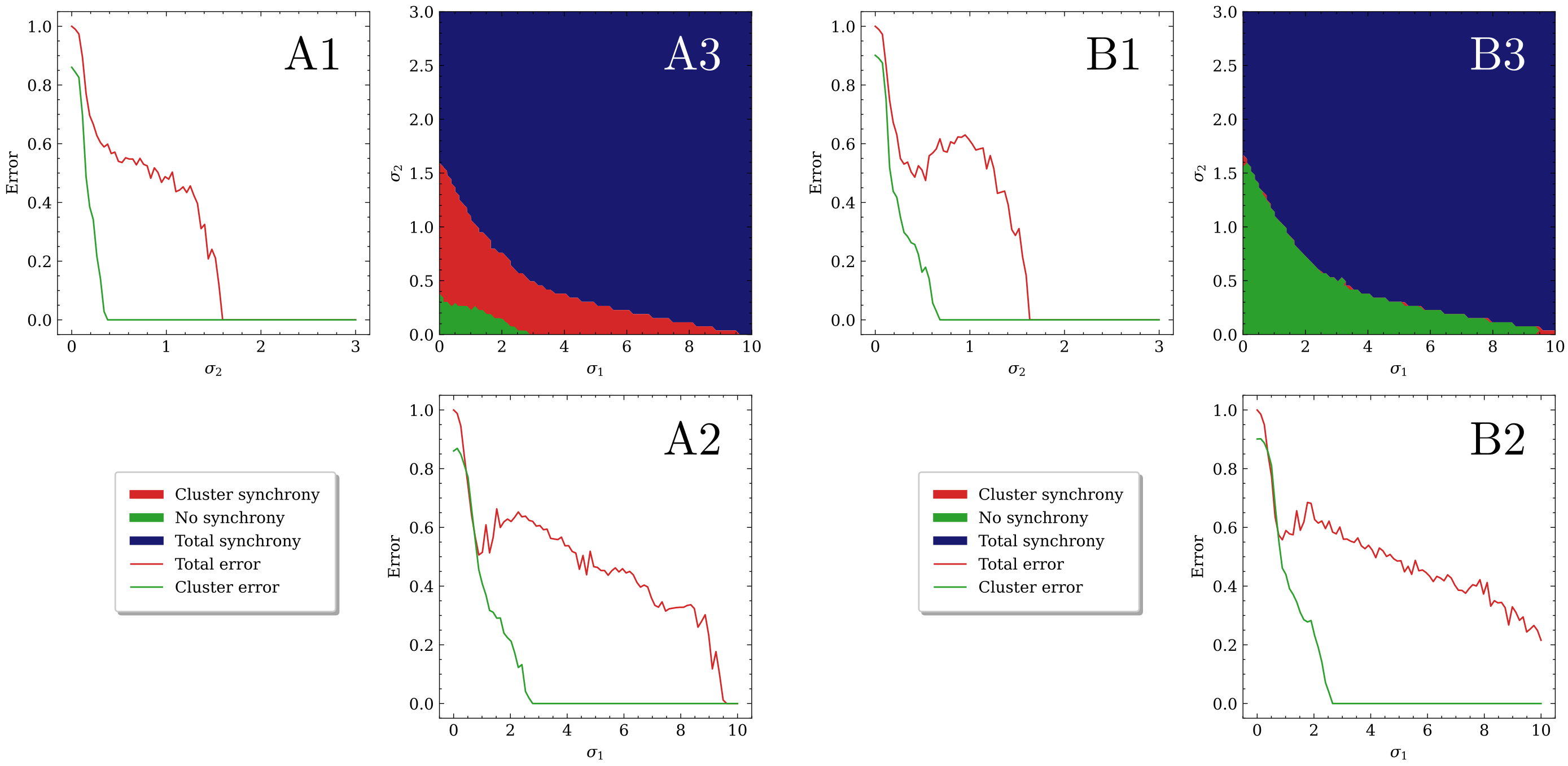}
    \caption{\textbf{Equitability and synchronisation on two multiplex networks.} Synchronization error \changes{Eq.~\eqref{eq:syncerror}} of coupled Lorenz oscillators in the two biplex networks shown in Fig.~\ref{fig:multilayer-condition-fulfilled-and-not-graphs} with respect to the cluster $\{7,8,9,10\}$. Panels A1 and A2 (respectively B1 and B2) show the cluster synchronization errors in $G_A$ (respectively $G_B$) when only layer 2, or layer 1, is active. Panels A3 and B3 show the $10^{-5}$ contours of the synchronization errors \changes{(i.e. the regions whose error is below or above $10^{-5}$)} for $(\sigma_1,\sigma_2)\in [0,10]\times[0,10]$ for networks $G_A$ and $G_B$, respectively. See the SI for full details.}
    \label{fig:multilayer-condition-fulfilled-and-not}
\end{figure*}

The results of the simulations are shown in Fig.~\ref{fig:multilayer-condition-fulfilled-and-not}. Panels A1 to A3 (respectively B1 to B3) show the total synchronisation errors on the biplex network $G_A$ (respectively $G_B$) as well as the cluster synchronisation error for the nodes $\{7,8,9,10\}$, which form a layer-by-layer equitable cluster in $G_A$ (but not in $G_B$). 
Panels A1, A2, B1 and B2 show that cluster synchronisation occurs on each layer separately (the cluster synchronisation error drops before the global one does) for both $G_A$ and $G_B$. However, only $G_A$ exhibits multilayer cluster synchronisation: panel A3 shows a significant region of coupling values $(\sigma_1,\sigma_2)$ (red area) for which the cluster $\{7,8,9,10\}$ synchronises while the network is not globally synchronised; this is however numerically negligible for $G_B$, as shown in panel B3, which suggests that the system undergoes a critical transition from no synchronisation to global synchronisation (explosive synchronisation). 

The general results in our paper fully explain the numerically observed behaviour: the nodes $\{7,8,9,10\}$ form a layer-by-layer equitable partition on $G_A$, but no such partition exists on $G_B$ (Fig.~\ref{fig:multilayer-condition-fulfilled-and-not-graphs}), exemplifying the prevalence of explosive synchronisation on multilayer (and higher-order) networks compared to standard (monolayer) networks. (See also the `Explosive synchronisation and equitability' section below.)

\subsection{Results for hypergraph dynamics}

Now we turn our attention to higher-order networks (hypergraphs), with dynamical interactions described by Eqs.~\eqref{eq:dynamical-system-general} and \eqref{eq:coupling-hypergraph}, and show that all our previous results hold in this setting. In this section, we state the key definitions and main results, with full details deferred to the SI. 

We first introduce the following notation, where $\cP$ is a partition of the node set, $C, C_1,\ldots, C_m \in \cP$, $i \in C$, and $1 \le m \le M$,
\begin{align}
    h_i^{C_1,\ldots,C_m} &= \sum_{j_1\in C_1} \ldots \sum_{j_1\in C_m}A^{(m)}_{i,j_1,\ldots,j_m}. \label{eq:hyper-structural-input-1}
\end{align}
We also define corresponding dynamical quantities $\bh_i^{C_1,\ldots,C_m}$, given a solution $\vec\bx=(\bx_1,\ldots,\bx_N)$, by adding $\bg^{(m)}(\bx_i,\bx_{j_1},\ldots,\bx_{j_m})$ to Eq.~\eqref{eq:hyper-structural-input-1} above. These quantities account for the total structural, respectively dynamical, input to node $i$ from all nodes in $C_1,\ldots,C_m$.

Unlike the 2-body (pairwise) case, we allow in general the `input' clusters $C_1,\ldots,C_m$ to be equal to $C$ or to each other since interactions are now not only internal ($i,j_1,\ldots,j_m \in C$) or external ($j_1,\ldots,j_m \not\in C$) but `mixed' in general. Dynamical equitability is still independent of the purely internal connectivity (due to the non-invasiveness of the coupling functions $\bg^{(m)}$) so we remove the case $C_1=\ldots=C_m=C$ in the definitions of equitability, below.  

We call a $\cP$-synchronised solution $\vec\bx$ \emph{dynamically $\cP$-equitable} if, for all $C \in \cP$, $i, j \in C$, we have 
\begin{equation}\label{eq:dynamical-equitability-hypergraph}
    \bh_i^\text{ext} = \bh_j^\text{ext},
\end{equation}
where $\bh_i^\text{ext} = \bh_i - \bh_i^\text{int} = \bh_i - \sum_{m=1}^M \sigma_m \, \bh_i^{C,\stackrel{m}{\ldots},C}$ with $C,\stackrel{m}{\ldots},C$ representing $C$ repeated $m$ times.
We call the partition $\cP$ \emph{externally equitable} if, for all $C \in \cP$, $i, j \in C$, we have 
\begin{equation}\label{eq:structural-equitability-hypergraph}
    h_i^{C_1,\ldots,C_m} = h_j^{C_1,\ldots,C_m},
\end{equation}
for all $C_1,\ldots, C_m \in \cP$ not all equal to $C$, and all $1 \le m \le M$. 

Similarly, we extend the definition of independent cluster synchronisation to multi-body interactions. We call a $\cP$-synchronised solution $\vec\bx$ \emph{$\cP$-independent} if, for every $C \in \cP$, the family of functions 
\begin{equation}\label{eq:linear-independence-hypergraph}
    \bg^{(m)}(\bx_C, \bx_{C_1}, \ldots, \bx_{C_m})
\end{equation}
where $1 \le m \le M$, and $C_1,\ldots, C_m \in \cP$ not all equal to $C$, is linearly independent over the time domain $T$. 

As in the multiplex case, we can show that cluster synchronisation implies dynamical equitability, and an equivalence between dynamical and external equitability for independent cluster synchronisation (see the SI for precise mathematical statements and proofs). \changes{In particular, we have that cluster synchronisation can only occur in one of these cases:
\begin{itemize}
    \item there is a linear dependence relation between the coupling functions on trajectories; or, if not, 
    \item the partition of the underlying interaction hypergraph must be externally equitable (Eq.~\eqref{eq:structural-equitability-hypergraph}). 
\end{itemize}}

\begin{figure} 
    \centering
    \includegraphics[width=1\linewidth]{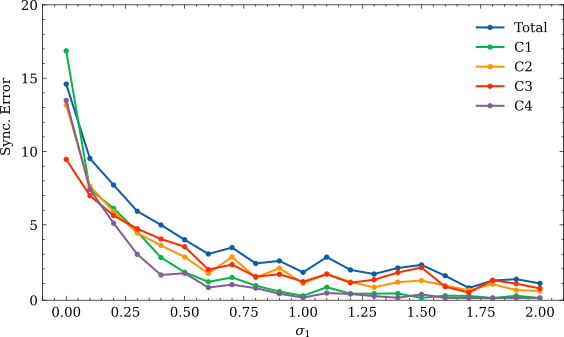}
    \caption{\textbf{Equitability and synchronisation on a hypergraph.} Synchronization error of coupled Lorenz oscillators for four cluster on a hypergraph (fully described in the SI) with 20 nodes and both pairwise and triadic interactions. The clusters satisfy only pairwise respectively triadic structural equitability (clusters $C_1$ respectively $C_2$), none (cluster $C_3$) or simultaneously both (cluster $C_4$). The coupling parameter $\sigma_2$ is fixed at 0.2 while $\sigma_1$ is increased. Only cluster $C_4$ synchronises before the whole system does, as shown by the synchronisation error, which agrees with our prediction, as the only simultaneously equitable cluster for all types of interactions. Cluster $C_1$ becomes almost synchronous due to the triadic interactions becoming small with respect to the pairwise interactions for which $C_1$ does have external equitability.
    }
    \label{fig:error-lowsigma2.png}
\end{figure}

To demonstrate our theoretical results numerically, we simulated Lorenz oscillators in a hypergraph with 20 nodes, four clusters, and both 2- and 3-body interactions (see the SI for a full description of the system). The clusters were chosen such that $C_1$ is (externally) equitable with respect to pairwise but not triadic interactions; cluster $C_2$ is equitable for triadic but not pairwise interactions; cluster $C_3$ satisfy neither pairwise nor triadic equitability; and cluster $C_4$ is the only one that simultaneously satisfy pairwise and triadic equitability. 
The results (Fig.~\ref{fig:error-lowsigma2.png}) show the mean square error of the four clusters for a fixed and small value of the triadic interaction coupling ($\sigma_2=0.2$), and varying pairwise interaction coupling ($\sigma_1$). We can see that only cluster $C_4$ synchronises before the whole system does. We also note that cluster $C_1$ becomes almost synchronous due to triadic interactions becoming small with respect to the pairwise interactions for which $C_1$ does satisfy external equitability.

Finally, we can also prove the reciprocal statement, that is, that the external equitability of a given partition implies the existence of cluster synchronised solutions, by `lifting' (repeating on each cluster) a solution of the quotient dynamics to the parent hypergraph. For completeness, we give below the definitions of quotient dynamics for multi-body interactions, with full details included in the SI. 

If $A$ is a square tensor (that is, all indices have the same dimension $N$, for example an adjacency tensor for a hypergraph with $N$ nodes) and $\cP$ is a partition of the set $\{1,\ldots,N\}$ into $K$ clusters, the \emph{quotient tensor} of $A$ with respect to $\cP$ is the square tensor $B$ defined by 
\begin{equation}
    B_{k_1,\ldots,k_m} = \frac{1}{|C_{k_1}|} \sum_{i_1 \in C_{k_1}, \ldots, i_m \in C_{k_m}} A_{i_1,\ldots, i_m},
\end{equation}
the average connectivity of nodes in $C_{k_2}, \ldots, C_{k_m}$ to a node in $C_{k_1}$. 
The \emph{quotient dynamics} of the dynamical system Eqs.~\eqref{eq:dynamical-system-general}, \eqref{eq:coupling-hypergraph} with respect to a partition $\cP$ is the system with dynamical units $\by_1,\ldots,\by_K$, where $K=|\cP|$, and differential equations
\begin{equation}
    \dot{\by}_i = \mathbf{f}(\by_i) + \bh_i(\by_1,\ldots,\by_K),
\end{equation}
where 
\begin{multline} \label{eq:quotient-dynamics-hypergraph}
 \bh_{i}(\by_1,\ldots,\by_K) = \sum_{m=1}^M\sigma_m\\ \sum_{j_1,\ldots,j_m=1}^K B^{(m)}_{i,j_1,\dotsc,j_m}\mathbf g^{(m)}(\by_{i}, \by_{j_1}, \dotsc, \by_{j_m})\:,
\end{multline}
and $B^{(m)}$ is the quotient tensor of $A^{(m)}$ with respect to $\cP$, for each $m$.

\changes{Similarly to the multiplex case, we show (in the SI) that $(\bx_1,\ldots,\bx_N)$ is a $\cP$-synchronised solution of the parent dynamics if and only if $(\by_1,\ldots,\by_K)$ is a solution of the quotient dynamics, where $\by_k=\bx_i$ whenever $i \in C_k \in \cP$. This means that any solution of the quotient system (with respect to an externally equitable partition) gives rise to a cluster synchronised solution (with respect to the clusters in the partition) of the original (parent) system.}

\startchanges
\subsection{General coupling and general equitability}
Our approach can be adapted to general coupling functions, that is, dropping the non-invasiveness condition $\bg(x,\ldots,x) = 0$ in Eqs.~\eqref{eq:coupling-multilayer} and \eqref{eq:coupling-hypergraph}, by substituting external equitability with general equitability. That is, we need to consider the internal cluster connectivity which is not necessarily zero when the cluster is synchronised in this general case. 

Our previous results can be extended to general coupling functions with the following changes from external to general equitability:
\begin{itemize}
    \item substitute \eqref{eq:weak-external-dynamical-equitability} and \eqref{eq:dynamical-equitability-hypergraph} by $\bh_i = \bh_j$;
    \item remove $C \neq C'$ before \eqref{eq:structural-equitability} and \eqref{eq:cluster-independence}; 
    \item remove `not all equal to $C$' after \eqref{eq:structural-equitability-hypergraph} and \eqref{eq:linear-independence-hypergraph}. 
\end{itemize}

With these changes, our results hold: for linear independent solutions, cluster synchronisation implies (general) equitability and vice-versa (see the SI for precise mathematical statements and derivations).   

Perhaps counter-intuitively at first, the generic case includes the non-invasive case but needs to be treated separately: if any of the coupling functions are non-invasive (any of the functions satisfy $\bg(\bx,\ldots,\bx) = 0$, the solutions are never linearly independent (the family of functions Eq.~\eqref{eq:cluster-independence} includes a zero function when $C=C'$) and our results do not apply (no linearly independent solution exists). If all functions are non-invasive (they all satisfy $\bg(\bx,\ldots,\bx) = 0$), our previous approach (first part of the paper) applies, by restricting to external equitability.

More generally, if we have a mixed system with non-invasive and general coupling functions, our results apply to the corresponding mixed notion of equitability: external equitability on layer $m$, respectively order $m$, if $g^{(m)}$ is non-invasive, otherwise general equitability, and adapting the notion of linear independence condition (remove $\bg^{(m)}(\bx_C,\bx_C)$ whenever $\bg^{(m)}$ is non-invasive). The same also applies to a general coupling function that becomes non-invasive on a particular synchronised trajectory ($\bg(\bx_C,\bx_C) = 0$ for some cluster $C$). 



\subsection{Explosive synchronisation and equitability}
Explosive synchronisation refers to the sudden transition from a non-synchronised to a fully synchronised state of the dynamical units. Our results guarantee that explosive synchronisation is the only possible generic (that is, linearly independent) synchronisation that can occur if the only equitable partitions are partitions into 1 or $N$ clusters (corresponding to global, respectively no synchronisation). Let us refer to this as \emph{generic explosive synchronisation}, that is, systems for which the only generic (linearly independent) solution is the globally synchronised one ($\bx_i=\bx_j$ for all $i, j$). 

In the non-invasive case, the partition into one cluster is always externally equitable (condition \eqref{eq:structural-equitability} is empty, as there is no external connectivity to consider) and, indeed, any solution of the internal dynamics $\dot\bx = \mathbf{f}(\bx)$ is a globally synchronised solution of the system, as explained earlier. Therefore, we have generic explosive synchronisation if and only if the underlying partition have no external equitable partitions except the partitions into one and into $N$ clusters --- a purely structural condition of the underlying interaction multiplex or hypergraph.  

In the general case, however, the partition into one cluster may not be equitable, as we now need to consider internal equitability. Indeed, the system admits a globally synchronised solution if and only if the partition into one cluster is equitable. This, in turn, is equivalent to a regularity condition on the underlying multiplex, or hypergraph, namely that all nodes have the same weighted in-degree on each layer, respectively for each order $m$ (see the SI for full details). 


\subsection{Finding equitable partitions}


Finding all equitable partition on an arbitrary graph (network) is in general a computationally hard problem \cite{lubiw1981some}, but there are some tractable cases. For example, the refinement procedure in the Weisfeiler-Lehmann (WL) algorithm \cite{huang2021short} finds the coarsest equitable partition in polynomial time --- any other equitable partition must be a refinement (obtained but further subpartitioning the clusters). Any other equitable partition must therefore be a refinement of the WL one. In particular, only nodes within the same WL cluster can synchronise (in a generic, that is, linearly independent, way).  
The WL refinement algorithm can be easily adapted to weighted networks, multiplexes, and hypergraphs, and to obtain the coarsest equitable refinement of a given partition $\cP$ (see SI for details). 


The case of external equitability is more subtle. Since the computation of all equitable partitions is computationally hard, so is the computation of all external equitable partitions. Any equitable partition, including the WL one, is also externally equitable, but there can be more (for example, the partition into one cluster is always externally equitable). The WL algorithm can be adapted to find the coarsest externally equitable partition containing a given partition $\cP$ (see the SI) but it cannot be used to find the coarsest partition overall: the partition into one cluster (whole network) is already externally equitable.  

It would be very interesting to find algorithms to find equitable and externally equitable partitions in the large but sparse graphs typically found in applications, but this is out of the scope of the present article. 

\finishchanges

\section{Conclusions}

\changes{In this article, we are concerned with very general necessary and sufficient conditions for the existence of cluster synchronised solutions on two very general systems of coupled dynamics (multiplex, and hypergraph).} We show that cluster synchronisation, with respect to a partition $\cP$, can only occur in either of these two cases:
\begin{itemize}
    \item there is a linear dependence relation between the coupling functions on trajectories (see Eqs.\eqref{eq:cluster-independence} and \eqref{eq:linear-independence-hypergraph}; or, if not, 
    \item $\cP$ is an equitable partition of the underlying interaction multiplex or hypergraph (Eq.~\eqref{eq:structural-equitability} respectively \eqref{eq:structural-equitability-hypergraph}). 
\end{itemize}
\changes{In other words, cluster synchronisation can occur without equitability (of the underlying cluster partition), but only if there are linear dependencies between the coupling functions.}
The first case (linear dependence) depends on the particular algebraic properties of the coupling functions, while the second one (equitability) is a structural property of the underlying pattern of interactions, and thus independent of the particular form of the coupling function.  

Moreover, in the linear independent, or equitable, case, we show that the correct notion of external equitability is simultaneous equitability per layer (multiplex case, Eq.~\eqref{eq:structural-equitability}) or per many-body interaction (hypergraph case, Eq.~\eqref{eq:structural-equitability-hypergraph}), explaining why cluster synchronisation is much harder to achieve in general in multi-layer and higher-order coupled dynamical systems. Our theoretical findings are illustrated with numerical simulations for both multiplex networks and hypergraphs. 

Although we cannot make any general claims about the stability of the cluster synchronised solutions, which will depend on the specific form of the coupling functions and values of the coupling strength parameters, our results provide strong restrictions about the groupings of nodes that can independently synchronise for any choice of coupling function, based solely on the equitability of the underlying interaction network, multiplex, or hypergraph. 

Finally, using the quotient dynamics, we also show that equitable solutions exists for any pre-determined clustering structure, paving the way for control applications in clustering engineering and control on arbitrary systems. 



\changes{There are limitations to our approach, for instance when there are linear dependencies between coupling functions either overall, or when restricted to some synchronised solutions: such generic solutions can exists without equitability of the underlying partition. (For a concrete example, inspired by the natural coupling in \cite{gambuzza2021stability}, see the SI.) Again, such solutions are not `generic enough' as they depend on linear dependencies of the coupling functions. }

Our work focuses on two general systems of dynamical interactions, each generalising standard pairwise (network) dynamics: multiplex (different `types' or layers of edges or pairwise interactions) and hypergraph (multi-body interactions). Of course, a even more general system would combine several layers of multi-body interactions --- the interested reader can easily adapt the results in our paper to such systems as needed.

Our results formalise and clarify the relationship between cluster synchronisation and equitability, but several important open questions remain. These include fast and exhaustive algorithms to find equitable partitions in arbitrary multiplexes and hypergraphs; the realisation and ordering problem, that is, which equitable partitions and in which order they synchronise as we increase the coupling strength parameters (see \cite{bayani2024transition} for the network case); the stability question, that is, finding general conditions that guarantee the stability and a synchronised solution, for instance from a quotient to a parent solution; and an extension to non-identical dynamical units such as general multi-layer networks and to other synchronisation types beyond identical synchronisation. \changes{We hope our article will stimulate work on these, and other, directions.}



\medskip 

\noindent\textbf{Acknowledgements.}
G.C.-A. is partially funded by the URJC under fellowship PREDOC-21-026-2164 and the INCIBE/URJC Agreement M3386/2024/0031/001 within the framework of the Recovery, Transformation and Resilience Plan funds of the European Union (Next Generation EU). C.I.d.G. acknowledges funding from the Bulgarian Ministry of Education and Science, under Project No.~BG-RRP-2.004-0006-C02. R.J.S.-G. has been partially supported by the Erlangen Hub: Mathematical Foundations of Intelligence grant EP/Y028872/1. 

\medskip 

\noindent\textbf{Author contribution.}
R.J.S.-G.,~S.B.~and C.I.d.G.~conceived the project. The theoretical framework was developed by R.J.S.-G.~and K.K. The numerical simulations were done by G.C.-A.,~K.K.~and C.I.d.G. All authors wrote and reviewed the final manuscript.

\smallskip 

\noindent\textbf{Data availability.} All data needed to replicate the findings is synthetic and can be reconstructed from the description in the Supplementary Information. 




\bibliographystyle{abbrv} 
\bibliography{bibliography} 


\onecolumngrid

\newpage

\begin{center}
    \Large Supplementary Information
\end{center}

\section{Theoretical results}
\subsection{Dynamical system and synchronisation}\label{sec:dynamical_system}
We consider the dynamical system introduced in the Main Text, that is, the system of $N \ge 1$ differential equations with $N$ dynamical variables $\bx_{1},\dots, \bx_N \colon T \to \mathbb{R}^D$, for some $T \subseteq \mathbb{R}$ open interval (\emph{time domain}) and $D \ge 1$ integer (\emph{dimension}), given by
\begin{equation}\label{eq:dynamical-system-general-SI}
    \dot{\bx}_i = \bbf(\bx_i) + \mathbf{h}_i(\bx_{1},\dots, \bx_N) \qquad i=1,\ldots,N
\end{equation}
with either $\mathbf{h}_i = \mathbf{h}_i^{multi}$ for all $i$ (called the \emph{multilayer case}), or $\mathbf{h}_i = \mathbf{h}_i^{hyper}$ for all $i$ (called the \emph{hypergraph case}), where 
\begin{align}
    \mathbf{h}_i^{multi}(\bx_{1},\dots, \bx_N) &= \sum_{m=1}^M \sigma_m \sum_{j=1}^N A^{(m)}_{i,j} \mathbf{g}^{(m)}(\bx_i, \bx_j), \text{  and} \label{eq:coupling-multilayer-SI}\\
    \mathbf{h}_{i}^{hyper}(\bx_{1},\dots, \bx_N) &= \sum_{m=1}^{M} \sigma_m \sum_{j_1,\ldots,j_m=1}^N A^{(m)}_{i,j_1,...,j_{m}} \, \mathbf{g}^{(m)}(\bx_i,\bx_{j_1}, \dots, \bx_{j_m}).
    \label{eq:coupling-hypergraph-SI}
\end{align}
Here, we assume the following:
\begin{itemize}

    \item $\bbf \colon \mathbb{R}^{D} \to \mathbb{R}^D$ is a smooth function;

    \item each $\mathbf{g}^{(m)}$ is a smooth function of the form $ \mathbf{g}^{(m)}\colon \mathbb{R}^{2 \times D} \to \mathbb{R}^D$ (multilayer case) or $\mathbf{g}^{(m)}\colon \mathbb{R}^{(M+1) \times D} \to \mathbb{R}^D$ (hypergraph case); 

    \item $\sigma_m > 0$ are non-negative real parameters (coupling strength for layer~$m$ or order~$m$);
    
    \item $A^{(m)}$ is the adjacency matrix for layer $m$, or the hypergraph adjacency tensor of order $m+1$; 

    \item the coupling functions are synchronisation non-invasive, that is, they satisfy\\ $\mathbf{g}^{(m)}(\bx,\ldots,\bx)~=~\mathbf{0}$ for all $m$, and any $\bx \in \mathbb{R}^D$. (For the general case, see Section \ref{sec:general-coupling}.)

\end{itemize}

The function $\mathbf{f}$ represents the internal dynamics at each node, the functions $\mathbf{g}^{(m)}$ the coupling dynamics on layer $m$ (multilayer case) or the $(m+1)$-body dynamics (hypergraph case), the matrix or tensor $A^{(m)}$ model the pairwise (multilayer case) or many-body (hypergraph case) interactions between the dynamical variables, and the coupling parameter $\sigma_m$ model the overall coupling strengths on layer $m$ (multilayer case) or on $m$-body interactions (hypergraph case). One important example is the Laplacian operator, $\mathbf{g}^{(m)}(\bx_i,\bx_j)=\bx_i-\bx_j$ --- this is indeed the only linear non-invasive operator in two variables up to a constant. 

\begin{remark}
    Mathematically, the coupling parameter can be absorbed into the adjacency matrix, respectively tensor. In practice, we use them to parametrise families of dynamical systems. For instance, in simulations, we will typically increase each $\sigma_m$ from zero, or very low values, to higher values to study the transition into (cluster and/or global) synchronisation of the system. 
\end{remark}

Note that, for $M=1$, both the multilayer and the hypergraph cases reduce to the network (monolayer) case, 
\begin{equation}\label{eq:dynamical-system-network-SI}
    \dot{\bx}_i = \bbf(\bx_i) + \sigma \sum_{j=1}^N a_{ij} \, \mathbf{g}(\bx_i, \bx_j).
\end{equation}

We can represent the interactions between the variables $\bx_1,\ldots,\bx_N$ using a multilayer network (multilayer case) or a hypergraph (hypergraph case). In the multilayer case, we define the \emph{interaction multilayer network} of this dynamical system as the weighted multilayer network with $M$ layers of $N$ vertices and inter-layer weighted adjacency matrices $A^{(m)}$ ($1 \le m \le M$). In the hypergraph case, the \emph{interaction hypergraph} is the weighted hypergraph with vertices $V = \{1,\ldots,N\}$ and weighted hyperedges $\{i_1,\ldots,i_m\}$ whenever $A^{(m)}_{i_1,\ldots,i_m} \neq 0$, which then becomes the weight of that hyperedge. 

\begin{remark}[Multiplex and multilayer dynamics]
    Strictly speaking, our multilayer model is a multiplex, as we have the same $N$ nodes (dynamical units) on each layer, that is, we have $M$ different `types' of pairwise interactions between the same $N$ nodes. In multilayer network, we have different nodes on each layer and both inter- and intra-layer pairwise interactions. The dynamics on a multilayer network with $N$ nodes organised into $M$ layers of $N_\alpha$ nodes each (so that $N=\sum_{\alpha=1}^M N_\alpha$) would have a coupling of the form
\begin{equation}
 \bh_i^{\mathrm{multilayer}} = \sum_{\beta=1}^M\sigma_{\alpha,\beta}\sum_{j \in V_\beta} A^{\alpha,\beta}_{i,j}\mathbf g^{\alpha, \beta}(\bx_i, \bx_j)\:,
\end{equation}
where $i$ is a node in the $\alpha$ layer and $V_\beta$ is the set of nodes in the $\beta$ layer. If all nodes (dynamical units) are identical (same internal dynamics given by the same function $\mathbf{f}$), this case can be seen as a multiplex network on $N$ nodes and $M^2$ layers $m=(\alpha,\beta)$ such that $A^{(m)}_{ij}$ is zero unless $i \in V_\alpha$ and $i \in V_\beta$. The generic multilayer case (different internal dynamics, that is, functions $\mathbf{f}^{(\alpha)}$, on each layer $\alpha$) is out of the scope of the present paper and part of our future work on equitablity and synchronisation.  
\end{remark}

A \emph{solution} $\bx = (\bx_1,\ldots,\bx_N)$ of the dynamical system \eqref{eq:dynamical-system-general} consists of differentiable functions $\bx_1, \ldots, \bx_N \colon T \to \mathbb{R}^m$ that satisfy the system of equations.  A \emph{synchronised solution} is a solution which satisfies $\bx_1 = \ldots = \bx_N$ (identical synchronisation). By the non-invasive property above, $\bx_s$ is a solution of the differential equation $\dot{\bx} = \bbf(\bx)$ if and only if $\bx_1 = \ldots = \bx_N = \bx_s$ is a (synchronised) solution of \eqref{eq:dynamical-system-general}. 

A subset $C$ of the node set $V = \{1,\ldots,N\}$ is called a \emph{cluster}. A solution $\bx_1, \ldots, \bx_N$ is $C$-\emph{synchronised}, or \emph{synchronised on $C$}, if $\bx_i = \bx_j$ for all $i, j \in C$. Note that if $C = \{i\}$ a singleton (a set with one element), every solution is $C$-synchronised. 
If $\mathcal{P}$ is a partition of the node set $V$ (a collection of pairwise disjoint clusters whose union is $V$), we call a solution $\bx = (\bx_1, \ldots, \bx_N)$ \emph{$\mathcal{P}$-synchronised} if it is $C$-synchronised for each $C \in \mathcal{P}$.

\subsection{Dynamical and structural equitability} \label{sec:dynamical-equitability}

In this section, we discuss dynamical and structural equitability, in more generality than in the Main Text. We focus on the multiplex case, and discuss the hypergraph case in Section \ref{sec:dynamics-hypergraphs}.

\smallskip

\noindent\textbf{Correspondence to terminology in the Main Text.} Dynamical equitability in the Main Text corresponds to weak dynamically equitability (WDE) in this SI. Similarly, external equitability corresponds to structural equitability per layer (SEL), and linear independence in cluster synchronisation to linear independence with respect to the families $\mathcal{F}(C)$. 

Consider the multilayer dynamical system given by Equations \eqref{eq:dynamical-system-general} and \eqref{eq:coupling-multilayer-SI}. Let $\bx$ be a solution, $C \subseteq V$ a cluster, $i \in V$ a node, and $m \in \{1,\ldots, M\}$ a layer. We extend the notation from the Main Text to 
\begin{align}
    \bh_i &= \sum_{m=1}^M \sigma_m \sum_{j=1}^N A^{(m)}_{i,j} \mathbf{g}^{(m)}(\bx_i, \bx_j), &
    h_i &= \sum_{m=1}^M \sigma_m \sum_{j=1}^N A^{(m)}_{i,j}, \label{row1}\\
    \bh_i^{C} &= \sum_{m=1}^M \sigma_m \sum_{j\in C} A^{(m)}_{i,j} \mathbf{g}^{(m)}(\bx_i, \bx_j), &
    h_i^{C} &= \sum_{m=1}^M \sigma_m \sum_{j\in C} A^{(m)}_{i,j},\label{row2}\\  
    \bh_i^{C,m} &= \sum_{j\in C} A^{(m)}_{i,j} \mathbf{g}^{(m)}(\bx_i, \bx_j), &
    h_i^{C',m} &= \sum_{j\in C} A^{(m)}_{i,j},\label{row3}
\end{align}
and note the equalities 
\begin{align}
    \bh_i  &= \sum_{C \in \cP} \bh_i^C,   
    & \bh_i^C  &= \sum_{m=1}^M \bh_i^{C,m}, 
    & h_i  &= \sum_{C \in \cP} h_i^C,
    &\text{and}& & h_i^C  &= \sum_{m=1}^M h_i^{C,m}.
\end{align}
The rows above \eqref{row1} to \eqref{row3} encode the dynamical (left) respectively structural (right) input to node $i$ from all other nodes (top row), all nodes on cluster $C$ from all layer (middle row), and all nodes on cluster $C$ and layer $m$ (bottom row). The quantities on the left column, $\bh_i$, $\bh_i^C$ and $\bh_i^{C,m}$, depend on the solution $\bx$ and thus are functions of $t \in T$, and note that $\bh_i$ is simply $\bh_i^\text{multi}$ in Eq.~\eqref{eq:coupling-multilayer-SI}. The quantifies on the right column depend only on the underlying interaction multiplex (they can be seen as total or partial node degrees in the multiplex) and thus independent on any solution or exact form of the coupling functions --- they only depend on the pattern and strength of interactions between dynamical units. 

Next, we define several notions of external equitability. In all of them, equitability refers to `equal input' for each node in a cluster $C \subseteq V$ from nodes outside the cluster (`external'). The different notions arise from whether the input refers to the overall contribution from all nodes outside the cluster (`weak') or from nodes on each cluster in a partition separately, or on each cluster and layer (`per layer'), and to either the dynamical input with respect to a solution (`dynamical') or to the structural input with respect to the pattern and strength of interactions between dynamical units (`structural'). In what follows, we drop the word `external', for simplicity. 
 
\begin{definition}\label{def:equitability-multiplex}
We call a cluster $C \subseteq V$ \emph{dynamically equitable} with respect to a solution $\bx$ if every pair of nodes $i, j \in C$ have the same dynamical input from the nodes outside $C$, that is, 
\begin{align}\label{eq:cluster-dynamically-equitable}
    \bh_i^{V \setminus C} &= \bh_j^{V \setminus C}.
\end{align}
We call a cluster $C \subseteq V$ \emph{structurally equitable} if every pair of nodes $i, j \in C$ have the same structural input from the nodes outside $C$, that is, 
\begin{align}\label{eq:cluster-structurally-equitable}
    h_i^{V \setminus C} &= h_j^{V \setminus C}.
\end{align}
Let $\cP$ be a partition of the node set and $\bx$ a solution of the dynamical system. We call the partition $\cP$ \emph{weakly dynamically equitable (WDE)} (with respect to $\bx$) if every cluster $C \in \cP$ is dynamically equitable (with respect to $\bx$). We call the partition $\cP$ \emph{weakly structurally equitable (WSE)} if every cluster $C \in \cP$ is structurally equitable.  
We call the partition $\cP$ \emph{dynamically equitable (DE)} (with respect to $\bx$) if for every $C, C' \in \cP$, $C \neq C'$, and every $i, j \in C$, 
\begin{equation}\label{eq:DE}
    \bh^{C'}_i = \bh^{C'}_j.
\end{equation}
We call the partition $\cP$ \emph{structurally equitable (SE)} if for every $C, C' \in \cP$, $C \neq C'$, and every $i, j \in C$, 
\begin{equation}\label{eq:SE}
    h^{C'}_i = h^{C'}_j.
\end{equation}
We call the partition $\cP$ \emph{dynamically equitable per layer (DEL)} (with respect to $\bx$) if for every $C, C' \in \cP$, $C \neq C'$, and every $i, j \in C$, 
\begin{equation}\label{eq:DEL}
    \bh^{C',m}_i = \bh^{C',m}_j.
\end{equation}
A partition $\cP$ of the node set is \emph{structurally equitable per layer (SEL)} if for every $C, C' \in \cP$, $C \neq C'$, and every $i, j \in C$, 
\begin{equation}\label{eq:SEL}
    h^{C',m}_i = h^{C',m}_j.
\end{equation}
\end{definition}

\begin{remark}\label{rmk:external}
    Note that these definitions (WDE, WSE, DE, SE, DEL and SEL) are `external', in the sense that we do not need to hold in the case $C=C'$. 
\end{remark}

Next, we show that strong equitability implies weak equitability, equitability per layer implies equitability, and structural equitability per layer implies dynamical equitability for synchronised solutions. This, proved in Theorem \ref{thm:structural-dynamical-equitability} below, can be represented schematically as
\begin{align}\label{eq:diagram-structural-dynamical-equitabilities}
    \xymatrix{
        DEL \ar@{=>}[r] & DE \ar@{=>}[r] & WDE \\
        SEL \ar@{=>}[r] \ar@{=>}[u]^{CS} & SE \ar@{=>}[r] & WSE
    }
\end{align}
where CS that stands for `cluster synchronisation' (that is, $\cP$-synchronisation), and the notation $A \stackrel{B}{\implies} C$ means `$A \text{ and } B$ implies $C$'. Note that structural concepts ($SEL$, $SE$, $WSE$) are with respect to a fixed partition $\cP$ of the node set, and the dynamical concepts ($DEL$, $DE$, $WDE$, $CS$) are with respect to the partition $\cP$ and a solution $\bx$ of the dynamical system.  

\begin{remark}
The concepts of SEL and SE, respectively DEL and DE, coincide if there is only one layer $m=1$, that is, in the monolayer, or network, case, but otherwise they are distinct. Similarly, we can give examples, both in the dynamical and the structural sense, of weakly equitable partitions that are not equitable. 
\end{remark}

\begin{remark}\label{rmk:SE-WSE-not-good}
    Neither SE nor WSE necessarily imply dynamical equitability in any of its forms  even under cluster synchronisation (it is not difficult to give counterexamples --- mathematically, it amounts to the condition $\sum_i a_i = \sum_i b_i$ not implying $\sum_i c_i a_i = \sum_i c_i b_i$ for more than two summands and constants $a_i$, $b_i$, $c_i$) and there does not seem to be a natural condition that would make that occur. In any case, cluster synchronisation relates to the strongest form of structural equitability, SEL, as shown later (Section \ref{sec:synchronisation-and-equitability}) so we consider SEL the appropriate notion of structural equitability in multiplex networks, as presented in the Main Text. 
\end{remark}

\begin{theorem}\label{thm:structural-dynamical-equitability}
    Consider the multilayer dynamical system given by Equations \eqref{eq:dynamical-system-general} and \eqref{eq:coupling-multilayer-SI}. Let $\cP$ be a partition of the node set and $\bx$ a solution of the dynamical system. Then, 
    \begin{itemize}
        \item[(1)] if $\cP$ is dynamically, respectively structurally, equitable, then each $C \in \cP$ is weakly dynamically, respectively structurally, equitable; 
        \item[(2)] if $\cP$ is dynamically, respectively structurally, equitable per layer, then it is dynamically, respectively structurally, equitable; 
        \item[(3)] if $\cP$ is structurally equitable per layer and $\bx$ is $\cP$-synchronised, then it is dynamically equitable per layer. 
    \end{itemize}
\end{theorem}

\begin{proof}
    (1) This follows directly from the definitions. If $i, j \in C \in \cP$ then 
    \begin{align}
        \bh_i^{V\setminus C} = \sum_{C \neq C' \in \cP} \bh_i^{C'}
        = \sum_{C \neq C' \in \cP} \bh_j^{C'} = \bh_i^{V\setminus C},
    \end{align}
    and, similarly, we can show that $h_i^{C\setminus V} = h_j^{C\setminus V}$.\\[5pt]
    (2) This follows directly from the definitions. If $i \in C \in \cP$ then 
    \begin{align}
        \bh_i^{C} = \sum_{m=1}^M \bh_i^{C,m} = \sum_{m=1}^M \bh_j^{C,m} = \bh_j^{C}
    \end{align}
    and, similarly, we can show that $h_i^{C} = h_j^{C}$.\\[5pt]
    (3) If $\bx$ is $\cP$-synchronised, we can write $\bx_C$ for the synchronised solution on a cluster $C \in \cP$, that is, $\bx_i = \bx_C$ for any $i \in C$. Then, given $i, j \in C$, $C \neq C' \in \cP$, and $m \in \{1,\ldots,M\}$,
    \begin{align}
        \bh_i^{C',m} = \sum_{k\in C'} A^{(m)}_{i,k} \mathbf{g}^{(m)}(\bx_i, \bx_k)
        = \sum_{k\in C'} A^{(m)}_{i,k} \mathbf{g}^{(m)}(\bx_C, \bx_{C'})
        = \mathbf{g}^{(m)}(\bx_C, \bx_{C'}) \, h_i^{C',m},
    \end{align}
    and similarly for $\bh_j^{C',m}$. Therefore, $h_i^{C',m} = h_j^{C',m}$ implies $\bh_i^{C',m} = \bh_j^{C',m}$ and we have dynamical equitability per layer. 
\end{proof}

\subsection{Linear independence over trajectories} \label{sec:linear-independence-trajectories}
Given functions $f_1(t), \ldots, f_K(t)$, $t \in T \subseteq \mathbb{R}$, the set $\mathcal{F} = \{f_1, \ldots, f_K\}$ is \emph{linearly independent over $T$} if the condition $a_1 f_1(t) + \ldots + a_K f_K(t) = 0$ for all $t \in T$, where $a_i \in \mathbb{R}$ are constants, implies $a_1=\ldots=a_K=0$. Linear independence can be checked for differentiable functions by using the Wronskian $W(t)$, a $K \times K$ determinant of the functions and their higher derivatives: if $W(t)$ is not identically zero over $T$, the functions must be linearly independent \cite{bostan2010wronskians}. A single function $f_1(t)$ is linearly independent if (and only if) it is not identically zero on $T$, and $f_1, \ldots, f_K$, $K \ge 2$ are linearly independent if no function $f_i$ is an exact linear combination of the rest, over $T$. 

\begin{definition}\label{def:families}
Let $\mathcal{P}$ be a partition of the node set, $C, C' \in \cP$ clusters with $C \neq C'$, and $\bx$ a $\mathcal{P}$-synchronised solution. We define the following sets of functions on $T$:
\begin{align}
    \mathcal{F}(C) &= \{ \mathbf{g}^{(m)}(\bx_C,\bx_{C'}) \mid C' \in \cP, 1 \le m \le M\},\\
    \mathcal{F}(C,C') &= \{ \mathbf{g}^{(m)}(\bx_C,\bx_{C'}) \mid 1 \le m \le M \},\\
    \mathcal{F}(C,C',m) &= \{ \mathbf{g}^{(m)}(\bx_C,\bx_{C'}) \}.
\end{align}
\end{definition}

Note that $\mathcal{F}(C,C',m)$ only has one element, and the inclusions 
\begin{equation}\label{eq:families-inclusions}
    \mathcal{F}(C,C',m) \subseteq \mathcal{F}(C,C') \subseteq \mathcal{F}(C)
\end{equation}
for any $C, C' \in \cP$, $1\le m \le M$. Any subset of a linearly independent set is linearly independent, which means that
\begin{itemize}
    \item if $\mathcal{F}(C)$ is linearly independent, so is $\mathcal{F}(C,C')$ for any $C' \in \cP$, $C' \neq C$;
    \item if $\mathcal{F}(C,C')$ is linearly independent, so is $\mathcal{F}(C,C',m)$ for any $m$.
\end{itemize}
Schematically, we can represent this as 
 \begin{align}\label{eq:diagram-linear-independence-concepts}
     \xymatrix{
        \mathcal{F}(C) \ar@{=>}[r] & \mathcal{F}(C, C') \ar@{=>}[r] & \mathcal{F}(C,C',m).
    }
\end{align} 

As in the Main Text, we define linear independence over the first family of sets.
\begin{definition}
    Let $\mathcal{P}$ be a partition of the node set and $\bx$ a $\mathcal{P}$-synchronised solution. We call $\bx$ \emph{linearly independent} (with respect to $\cP$) if $\mathcal{F}(C)$ is linearly independent for each $C \in \cP$. 
\end{definition}

The following result clarifies the relation between linear independence and different notions of equitability. 

\begin{theorem}\label{thm:equitability-and-linear-independence}
    Consider the multilayer dynamical system given by Equations \eqref{eq:dynamical-system-general} and \eqref{eq:coupling-multilayer-SI}. Let $\cP$ be a partition of the node set and $\bx$ a $\cP$-synchronised solution of the dynamical system.
    \begin{itemize}
        \item[(1)] Suppose that $\mathcal{F}(C,C',m)$ is linearly independent (equivalently, $\bg^{(m)}(x_C,x_{C'})$ is not identically zero over $T$) for all $C, C' \in \cP$, $C \neq C'$ and $1\le m \le M$. Then $\cP$ is DEL with respect to $\bx$ if and only if $\cP$ is SEL.
        \item[(2)] Suppose that $\mathcal{F}(C,C')$ is linearly independent for all $C, C' \in \cP$, $C \neq C'$. Then $\cP$ is DE with respect to $\bx$ if and only if $\cP$ is SEL.
        \item[(3)] Suppose that $\mathcal{F}(C)$ is linearly independent for all $C \in \cP$. Then $\cP$ is WDE with respect to $\bx$ if and only if $\cP$ is SEL.
    \end{itemize}
\end{theorem}

The results of this theorem can be represented schematically as
\begin{align}\label{diagram:equitability-and-linear-independence}
    \xymatrix{
        DEL 
        \ar@<-0.6ex>@{=>}[r]
        \ar@{<=>}[ddr]_{\mathcal{F}(C,C',m)\,} 
        & 
        DE 
        \ar@<-0.6ex>@{=>}[r] 
        \ar@<-0.6ex>@{=>}[l]_-{\mathcal{F}(C,C')} 
        \ar@{<=>}[dd]|{\mathcal{F}(C,C')} 
        & 
        WDE 
        \ar@<-0.6ex>@{=>}[l]_-{\mathcal{F}(C)} 
        \ar@{<=>}[ddl]^{\mathcal{F}(C)} \\
        \\
        & SEL &
    }
    \raisebox{-1.5cm}{\hspace{0.5cm} (CS)}
\end{align}
where we assume cluster synchronisation (CS), we have used $DEL \implies DE \implies WDE$ from the previous theorem, and the right-to-left arrows are a consequence of the other arrows and \eqref{eq:diagram-linear-independence-concepts}. 
In particular, if $\bx$ is $\cP$-synchronised and linearly independent ($\mathcal{F}(C)$ is linearly independent for each $C \in \cP$), then all four notions are SEL, DEL, DE, and WDE are equivalent.

\begin{proof}
    We will write $\bx_C$ for the synchronised solutions on a cluster $C \in \cP$, that is, $\bx_C =\bx_i$ for any $i \in C$.\\[5pt]
    (1) Let $C, C' \in \cP$, $C \neq C'$, $i, j \in C$, $m \in \{1,\ldots, M\}$, and $t \in T$ time domain. We have 
    \begin{gather*}
        \mathbf{h}_i^{C',m}(t) = \mathbf{h}_j^{C',m}(t)\iff\\
        \sum_{k \in C'} A^{(m)}_{i,k} \mathbf{g}^{(m)}(\bx_i(t),\bx_k(t)) = 
        \sum_{k \in C'} A^{(m)}_{j,k} \mathbf{g}^{(m)}(\bx_j(t),\bx_k(t)) \iff \\
        \mathbf{g}^{(m)}(\bx_C(t),\bx_{C'}(t)) \sum_{k \in C'} \left( A^{(m)}_{i,k} - A^{(m)}_{j,k} \right) = 0 \iff \\
        \mathbf{g}^{(m)}(\bx_C(t),\bx_{C'}(t)) \left( h_i^{C',m} - h_j^{C',m} \right) = 0
    \end{gather*}
    with each equation satisfied for all $t \in T$. By the linear independence hypothesis, this is equivalent to
    \begin{gather*}
        h_i^{C,m} - h_j^{C,m} = 0 \text{ for all $m$}
        \iff h_i^{C,m} = h_j^{C,m} \text{ for all $m$}.
    \end{gather*}
    All in all, we have DEL holds if and only if SEL holds. We note that one implication (SEL implies DEL) does not need the linear independence hypothesis, just $\cP$-synchronisation, which we knew from Theorem \ref{thm:structural-dynamical-equitability} or diagram \ref{eq:diagram-structural-dynamical-equitabilities}. \\[5pt]
    (2) The proof is very similar, and sketched below (we drop the time parameter $t$), with the same notation as in (1): 
    \begin{gather*}
        \mathbf{h}_i^{C'} = \mathbf{h}_j^{C'} \iff
        \\
        \sum_{m=1}^M \sigma_m \sum_{k \in C'} A^{(m)}_{i,k} \mathbf{g}^{(m)}(\bx_i,\bx_k) = 
        \sum_{m=1}^M \sigma_m \sum_{k \in C'} A^{(m)}_{j,k} \mathbf{g}^{(m)}(\bx_j,\bx_k) \iff 
        \\
        \sum_{m=1}^M \sigma_m \mathbf{g}^{(m)}(\bx_C,\bx_{C'}) \sum_{k \in C'} \left( A^{(m)}_{i,k} - A^{(m)}_{j,k} \right) = 0 \iff
        \\
        \sum_{m=1}^M \sigma_m \mathbf{g}^{(m)}(\bx_C,\bx_{C'}) \left( h_i^{C',m} - h_j^{C',m} \right) = 0 \stackrel{\mathcal{F}(C,C')}{\iff}
        \\
        \sigma_m \left( h_i^{C,m} - h_j^{C,m} \right) = 0 \stackrel{\sigma_m \neq 0} \iff
        \\
        h_i^{C',m} = h_j^{C',m}.
    \end{gather*}
    (3) Analogously to (1) and (2), and with the same notation, we have 
    \begin{gather*}
        \mathbf{h}_i^{V\setminus C} = \mathbf{h}_j^{V\setminus C} \iff
        \\
        \sum_{m=1}^M \sigma_m \sum_{k \not\in C} A^{(m)}_{i,k} \mathbf{g}^{(m)}(\bx_i,\bx_k) = 
        \sum_{m=1}^M \sigma_m \sum_{k \not\in C} A^{(m)}_{j,k} \mathbf{g}^{(m)}(\bx_j,\bx_k) \iff 
        \\
        \sum_{m=1}^M \sigma_m \sum_{C' \in \cP, C' \neq C} \sum_{k \in C'} \mathbf{g}^{(m)}(\bx_C,\bx_{C'}) \left( A^{(m)}_{i,k} - A^{(m)}_{j,k} \right) = 0 \iff
        \\
        \sum_{m=1}^M \sum_{C' \in \cP, C' \neq C} \sigma_m \left( h_i^{C',m} - h_j^{C',m} \right) \mathbf{g}^{(m)}(\bx_C,\bx_{C'}) = 0 \stackrel{\mathcal{F}(C)}{\iff}
        \\
        \sigma_m \left( h_i^{C,m} - h_j^{C,m} \right) = 0 \stackrel{\sigma_m \neq 0} \iff
        \\
        h_i^{C',m} = h_j^{C',m}. \qedhere
    \end{gather*}
\end{proof}

Now that we have explored the different notions of equitability and their relationships, we move on to study cluster synchronisation in the context of equitability.

\subsection{Synchronisation and equitability}
\label{sec:synchronisation-and-equitability}
We first show that equitability is a necessary condition for cluster synchronisation, in the following sense. 

\begin{theorem}\label{thm:cluster-synchronisation-implies-equitability-multilayer}
Consider the multilayer dynamical system given by Equations \eqref{eq:dynamical-system-general} and \eqref{eq:coupling-multilayer-SI}. Let $\cP$ be a partition of the node set $\{1,\ldots, N\}$ and $\bx$ a $\cP$-synchronised solution. Then, 
\begin{itemize}
    \item[(1)] the partition $\cP$ is WDE with respect to $\bx$;
    \item[(2)] if $\bx$ is linearly independent ($\mathcal{F}(C)$ is linearly independent for all $C \in \cP$), then $\cP$ is SEL. 
\end{itemize}
\end{theorem}

\begin{proof}
    (1) Let $C \in \cP$, $i, j \in C$. Since $\bx=(\bx_1,\ldots,\bx_N)$ is a solution  the dynamical system, 
    \begin{equation}
        \dot{\bx}_i = \mathbf{f}(\bx_i) + \bh_i(\bx_1,\ldots,\bx_N) = \mathbf{f}(\bx_i) + \bh^C_i + \bh^{V \setminus C}_i, 
    \end{equation}
    and similarly for $j$. Since $\bx$ is $\cP$-synchronised, $\bh^C_i = \bh^C_j = 0$ (as the coupling functions are non-invasive), and $\bx_i = \bx_j$. In particular, 
    \begin{align}
        \dot{\bx}_i = \dot{\bx}_j \implies \bh_i^{V \setminus C} = \bh_j^{V \setminus C} 
    \end{align}
    and $\cP$ is WDE with respect to $\bx$ by definition.\\[5pt]
    (2) By (1), we have WDE, which, under the linear independence hypothesis, is equivalent to SEL (and DEL, and DE) by Theorem \ref{thm:equitability-and-linear-independence}; see diagram \ref{diagram:equitability-and-linear-independence}.
\end{proof}

Next, we show the reciprocal, that is, that equitability is also a sufficient condition, using the quotient dynamics. 

As in the Main Text, we define the \emph{quotient} of the dynamical system \eqref{eq:dynamical-system-general}, \eqref{eq:coupling-multilayer-SI} with respect to a partition $\cP$ of the node set $\{1,\ldots,N\}$ into $K = |\cP|$ clusters as the dynamical system with dynamical units $\by_1,\ldots, \by_K$ and $K$ differential equations 
\begin{align}
    \label{eq:quotient-dynamics-1}
    \dot{\by}_i & = \bbf(\by_i) + \bh_i(\by_1,\ldots,\by_K), \quad \text{ where}\\
    \label{eq:quotient-dynamics-2}
    \bh_i(\by_1,\ldots, \by_K) &= \sum_{m=1}^M \sigma_m \sum_{j=1}^K B^{(m)}_{i,j}\, \bg^{(m)}(\by_i,\by_j) \quad \text{ and }\\
    \label{eq:quotient-dynamics-3}
    B^{(m)}_{ij} &= \frac{1}{|C_i|} \sum_{k \in C_i, l \in C_j} A^{(m)}_{kl},
\end{align}
the quotient matrix of $A^{(m)}$ with respect to $\cP$. 

\begin{theorem}\label{thm:quotient-multilayer}
    Consider the multilayer dynamical system given by Equations \eqref{eq:dynamical-system-general} and \eqref{eq:coupling-multilayer-SI}, which we refer to as the parent dynamical system, a partition of the node set $\cP$, and the associated quotient dynamical system given by Equations \eqref{eq:quotient-dynamics-1} to \eqref{eq:quotient-dynamics-3}. Suppose that $\cP$ is SEL. Then,  $\bx=(\bx_1,\ldots,\bx_N)$ is a $\cP$-synchronised solution of the parent dynamical system if and only if $\by=(\by_1,\ldots,\by_K)$ is a solution of the quotient dynamical system, where $\by_k = \bx_i$ whenever $i \in C_k$. 
\end{theorem}

\begin{proof}
    Note that the condition $\by_k = \bx_i$ whenever $i \in C_k$ uniquely defines $\by$ from $\bx$ (since $\bx$ is $\cP$-synchronised), and $\bx$ from $\by$ (since $\cP$ is a partition of the node set).
    Using the equitability condition (SEL), we can write 
    \begin{align}\label{eq:quotient-matrix-equitability}
        B_{kl}^{(m)} = 
        \frac{1}{|C_k|} \sum_{i \in C_k, j \in C_l} A^{(m)}_{ij} = 
        \frac{1}{|C_k|} \sum_{i \in C_k} \overbrace{\sum_{j \in C_l} A^{(m)}_{ij}}^{=h_i^{C_l,m}} \stackrel{\ref{eq:SEL}}{=} 
        \frac{1}{|C_k|} |C_k| h_i^{C_l} = 
        h_i^{C_l} =
        \sum_{j \in C_l} A^{(m)}_{ij},
    \end{align}
    for any choice $i \in C_k$ and any $k \neq l$. 
    Suppose that $\bx=(\bx_1,\ldots,\bx_N)$ and $\by=(\by_1,\ldots,\by_K)$ satisfy $\by_k = \bx_i$ whenever $i \in C_k$. Let $k \in \{1,\ldots, K\}$ and $i \in C_k$. Then 
    \begin{align}
        \bh_k(\by_1,\ldots,\by_K) 
        &= \sum_{m=1}^N \sigma_m \sum_{l=1}^K B^{(m)}_{kl} \bg^{(m)}(\by_k,\by_l)\\ 
        &\stackrel{\eqref{eq:quotient-matrix-equitability}}{=} \sum_{m=1}^N \sigma_m \sum_{l=1}^K \sum_{j \in C_l} A^{(m)}_{ij} \bg^{(m)}(\by_k,\by_l)\\
        &= \sum_{m=1}^N \sigma_m \sum_{l=1}^K \sum_{j \in C_l} A^{(m)}_{ij} \bg^{(m)}(\bx_i,\bx_j)\\
        &= \sum_{m=1}^N \sigma_m \sum_{j=1}^N A^{(m)}_{ij} \bg^{(m)}(\bx_i,\bx_j) = \bh_i(\bx_1,\ldots, \bx_N). \label{eq:quotient-last}
    \end{align}
    (Note that the second equality follows from \eqref{eq:quotient-matrix-equitability} for $k \neq l$ and from the non-invasiveness of $\bg^{(m)}$ for $k=l$.)
    All in all, if $i \in C_k$, the differential equation for $\bx_i$ in the parent dynamical system is the same as the differential equation for $\by_k$ in the quotient dyanamical system. This completes the proof. 
\end{proof}

\subsection{Synchronisation and equitability for hypergraphs} \label{sec:dynamics-hypergraphs}

We now discuss the hypergraph case. Consider the hypergraph dynamical system given by Equations \eqref{eq:dynamical-system-general} and \eqref{eq:coupling-hypergraph-SI}. Let $\cP$ be a partition of the node set $\{1,\ldots,N\}$, $1 \le m \le M$, $C, C_1,\ldots, C_m \in \cP$, and $i \in C$. Consider the following notation, which extends the notation in the Main Text, 
\begin{align}
    h_i^{C_1,\ldots,C_m} &= \sum_{j_1\in C_1} \ldots \sum_{j_1\in C_m}A^{(m)}_{i,j_1,\ldots,j_m}, \label{eq:hyper-structural-input-1-SI}\\
    h_i^{(m)} &= \sum_{C_1, \ldots, C_m \in \cP} h_i^{C_1,\ldots,C_m}, \label{eq:hyper-structural-input-2-SI}\\
    h_i &= \sum_{m=1}^M \sigma_m h_i^{(m)} \label{eq:hyper-structural-input-3-SI}.
\end{align}
and corresponding dynamical quantities $\bh_i^{C_1,\ldots,C_m}$, $\bh_i^{(m)}$ and $\bh_i$,  given a solution $\bx = (\bx_1,\ldots, \bx_N)$, by adding $\bg^{(m)}(\bx_i,\bx_{j_1},\ldots,\bx_{j_m})$ to Eq.~\eqref{eq:hyper-structural-input-1-SI} above. 

\begin{definition}\label{def:equitability-hypergraphs}
    We call the partition $\cP$ \emph{dynamically equitable per layer (DEL)}, respectively \emph{dynamically equitable (DE)}, respectively \emph{weakly dynamically equitable (WDE)} with respect to a solution $\bx$, if, for each $C \in \cP$ and $i, j \in C$, we have 
    \begin{align}
        \bh_i^{C_1,\ldots,C_m} &= \bh_j^{C_1,\ldots,C_m}, \text{ respectively}\\
        \bh_i^{(m)} &= \bh_j^{(m)}, \text{ respectively}\\
        \bh_i^{\text{ext}} &= \bh_j^{\text{ext}}, \label{eq:external-hypergraph}
    \end{align}
    where each equation must hold for all $C_1,\ldots, C_m \in \cP$ not all equal to $C$, and all $1 \le m \le M$, and we write $\bh_i^\text{ext} = \bh_i - \sum_{m=1}^M \sigma_m \, \bh_i^{C,\stackrel{m}{\ldots},C}$ where $C,\stackrel{m}{\ldots},C$ represents $C$ repeated $m$ times. 
\end{definition}

\begin{remark}
    We keep a similar terminology to the multiplex case, even though it does not make much sense to talk about `layers' in this case but rather about `multi-body interactions'. 
\end{remark}

\begin{remark}\label{rmk:external-hypergraph}
    This is still an `external' notion of equitability, as we exclude the case $C_1=\ldots=C_m=C$, as argued in the Main Text.
\end{remark}

\begin{definition}\label{def:SEL-hypergraph}
    We call a partition $\cP$ \emph{structurally equitable per layer} (SEL) if, for each $C \in \cP$ and $i,j \in C$, we have 
    \begin{equation}\label{eq:hypergraph-SEL}
        h_i^{C_1,\ldots,C_m} = h_j^{C_1,\ldots,C_m}
    \end{equation}
    for all $C_1\ldots,C_m \in \cP$ not all equal to $C$, and all $1 \le m \le M$.
\end{definition}

\begin{remark}
    As in Section \ref{sec:dynamical-equitability}, we could have defined structural equitability (SE) and weak structural equitability (WSE) and shown $SEL \implies SE \implies WSE$. However, as in the multilayer case, SEL is the appropriate notion of structural equitability, for analogous reasons (see Remark \ref{rmk:SE-WSE-not-good}). 
\end{remark}

As in the multilayer case, we can show that strong equitability implies weak equitability, equitability per layer implies equitability, and structural equitability per layer implies dynamical equitability for synchronised solutions. 

\begin{theorem}\label{thm:structural-dynamical-equitability-hypergraph}
    Consider the hypergraph dynamical system given by Equations \eqref{eq:dynamical-system-general} and \eqref{eq:coupling-hypergraph-SI}. Let $\cP$ be a partition of the node set and $\bx$ a solution of the dynamical system. Then, 
    \begin{itemize}
        \item[(1)] if $\cP$ is DE, then it is WDE;
        \item[(2)] if $\cP$ is DEL, then it is DE; 
        \item[(3)] if $\cP$ is SEL and $\bx$ is $\cP$-synchronised, then $\cP$ is DEL. 
    \end{itemize}
\end{theorem}

\begin{proof}
    (1) and (2) follow directly from the definitions, namely Eqs.~\eqref{eq:hyper-structural-input-2-SI} and \eqref{eq:hyper-structural-input-3-SI}.\\[5pt]
    (3) If $\bx$ is $\cP$-synchronised, we can write $\bx_C$ for the synchronised solution on a cluster $C \in \cP$, that is, $\bx_i = \bx_C$ for any $i \in C$. Let $C \in \cP$, $i, j \in C$, $m \in \{1,\ldots,M\}$, and $C_1,\ldots,C_m \in \cP$. Then 
    \begin{align}
        \bh_i^{C_1,\ldots,C_m} 
            &= \sum_{j_1 \in C_1, \ldots, j_m \in C_m} A^{(m)}_{i,j_1,\ldots,j_m} \mathbf{g}^{(m)}(\bx_i, \bx_{j_1}, \ldots, \bx_{j_m})\\
            &= \sum_{j_1 \in C_1, \ldots, j_m \in C_m} A^{(m)}_{i,j_1,\ldots,j_m} \mathbf{g}^{(m)}(\bx_C, \bx_{C_1}, \ldots, \bx_{C_m})\\
            &= \mathbf{g}^{(m)}(\bx_C, \bx_{C_1}, \ldots, \bx_{C_m}) \sum_{j_1 \in C_1, \ldots, j_m \in C_m} A^{(m)}_{i,j_1,\ldots,j_m} \\
            & = \mathbf{g}^{(m)}(\bx_C, \bx_{C_1}, \ldots, \bx_{C_m}) \, h_i^{C_1,\ldots,C_m},
    \end{align}
    and similarly for $\bh_j^{C_1,\ldots,C_m}$. By hypothesis, $h_i^{C_1,\ldots,C_m} = h_j^{C_1,\ldots,C_m}$ and hence, by the above, $
    \bh_i^{C_1,\ldots,C_m} = \bh_j^{C_1,\ldots,C_m}$. 
\end{proof}
The results of the theorem can be represented schematically as follows,
\begin{align}\label{eq:diagram-structural-dynamical-equitabilities-hypergraph}
    \xymatrix{
        DEL \ar@{=>}[r] & DE \ar@{=>}[r] & WDE \\
        SEL \ar@{=>}[u]^{CS}
    }
\end{align}
with the same notational conventions as in \eqref{eq:diagram-structural-dynamical-equitabilities}. 

\begin{remark}
The concepts of DEL and DE coincide in the monolayer, or network, case ($m=1$), but otherwise they are distinct. Similarly, we can give examples of WDE partitions that are not DE. 
\end{remark}

Next, we define the corresponding notions of linear independence in the hypergraph case.

\begin{definition}\label{def:families-hypergraph}
Let $\cP$ be a partition of the node set, $C, C_1\ldots, C_m \in \cP$, and $\bx$ a $\cP$-synchronised solution. We define the following sets of functions on $T$:
\begin{align*}
    &\mathcal{F}(C) = \left \{ \bg^{(m)}(\bx_C,\bx_{C_1},\ldots,\bx_{C_m}) \mid 1\le m \le M, C_1,\ldots, C_m \in \cP \text{ not all equal to $C$} \right\},\\
    &\mathcal{F}(C, m) = \left \{ \bg^{(m)}(\bx_C,\bx_{C_1},\ldots,\bx_{C_m}) \mid C_1,\ldots, C_m \in \cP \text{ not all equal to $C$} \right\},\\
    &\mathcal{F}(C, C_1, \ldots, C_m) = \left \{ \bg^{(m)}(\bx_C,\bx_{C_1},\ldots,\bx_{C_m}) \right\}.
\end{align*}

\end{definition}
Note that $\mathcal{F}(C, C_1, \ldots, C_m)$ has only one element, and the inclusions 
\begin{equation}
    \mathcal{F}(C, C_1, \ldots, C_m) \subseteq \mathcal{F}(C,m) \subseteq \mathcal{F}(C)
\end{equation}
for any $C, C_1, \ldots, C_m \in \cP$, $1\le m \le M$. Any subset of a linearly independent set is linearly independent, which means that
\begin{itemize}
    \item if $\mathcal{F}(C)$ is linearly independent, so is $\mathcal{F}(C,m)$ for any $1 \le m \le M$;
    \item if $\mathcal{F}(C,m)$ is linearly independent, so is $\mathcal{F}(C, C_1, \ldots, C_m)$ for any $C_1, \ldots, C_m \in \cP$.
\end{itemize}
Schematically, we can represent this as 
 \begin{align}\label{eq:diagram-linear-independence-concepts-hypergraph}
     \xymatrix{
        \mathcal{F}(C) \ar@{=>}[r] & \mathcal{F}(C, m) \ar@{=>}[r] & \mathcal{F}(C, C_1, \ldots, C_m).
    }
\end{align} 

As in the Main Text, we define linear independence over the first family of sets.
\begin{definition}
    Let $\mathcal{P}$ be a partition of the node set and $\bx$ a $\mathcal{P}$-synchronised solution. We call $\bx$ \emph{linearly independent} (with respect to $\cP$) if $\mathcal{F}(C)$ is linearly independent for each $C \in \cP$. 
\end{definition}

The following result clarifies the relation between linear independence and different notions of equitability in the hypergraph case. It is analogous to Theorem \ref{thm:equitability-and-linear-independence} and the proof (omitted) is very similar. 

\begin{theorem}\label{thm:equitability-and-linear-independence-hypergraph}
    Consider the hypergraph dynamical system given by Equations \eqref{eq:dynamical-system-general} and \eqref{eq:coupling-hypergraph-SI}. Let $\cP$ be a partition of the node set and $\bx$ a $\cP$-synchronised solution of the dynamical system.
    \begin{itemize}
        \item[(1)] Suppose that $\mathcal{F}(C,C_1,\ldots,C_m)$ is linearly independent (equivalently, the coupling function $\bg^{(m)}(\bx_C,\bx_{C_1},\ldots,\bx_{C_m})$ is not identically zero over $T$) for all $C \in \cP$, $C_1,\ldots, C_m \in \cP$ not all equal to $C$, and $1\le m \le M$. Then $\cP$ is DEL with respect to $\bx$ if and only if $\cP$ is SEL.
        \item[(2)] Suppose that $\mathcal{F}(C,m)$ is linearly independent for all $C \in \cP$, and $1\le m \le M$. Then $\cP$ is DE with respect to $\bx$ if and only if $\cP$ is SEL.
        \item[(3)] Suppose that $\mathcal{F}(C)$ is linearly independent for all $C \in \cP$. Then $\cP$ is WDE with respect to $\bx$ if and only if $\cP$ is SEL.
    \end{itemize}
\end{theorem}

The results in this theorem can be represented schematically as
\begin{align}\label{diagram:equitability-and-linear-independence-hypergraph}
    \xymatrix{
        DEL 
        \ar@<-0.6ex>@{=>}[r]
        \ar@{<=>}[ddr]_{\mathcal{F}(C,C_1,\ldots,C_m)\,} 
        & 
        DE 
        \ar@<-0.6ex>@{=>}[r] 
        \ar@<-0.6ex>@{=>}[l]_-{\mathcal{F}(C,m)} 
        \ar@{<=>}[dd]|{\mathcal{F}(C,m)} 
        & 
        WDE 
        \ar@<-0.6ex>@{=>}[l]_-{\mathcal{F}(C)} 
        \ar@{<=>}[ddl]^{\mathcal{F}(C)} \\
        \\
        & SEL &
    }
    \raisebox{-1.5cm}{\hspace{0.5cm} (CS)}
\end{align}
where we assume cluster synchronisation (CS), we have used $DEL \implies DE \implies WDE$ from the previous theorem, Theorem \ref{thm:structural-dynamical-equitability-hypergraph}, and the right-to-left arrows are a consequence of the other arrows and \eqref{eq:diagram-linear-independence-concepts-hypergraph}. 
In particular, if $\bx$ is $\cP$-synchronised and linearly independent ($\mathcal{F}(C)$ is linearly independent for each $C \in \cP$), then all four notions are SEL, DEL, DE, and WDE are equivalent.

Finally, we related cluster synchronisation and stability in the next two results, which adapt Theorems \ref{thm:cluster-synchronisation-implies-equitability-multilayer} and \ref{thm:quotient-multilayer} to the hypergraph case.

\begin{theorem}\label{thm:cluster-synchronisation-implies-equitability-hypergraph}
Consider the dynamical system \eqref{eq:dynamical-system-general} with \eqref{eq:coupling-hypergraph-SI}. Let $\cP$ be a partition of the node set $\{1,\ldots, N\}$ and $\bx$ a $\cP$-synchronised solution. Then, 
\begin{itemize}
    \item[(1)] the partition $\cP$ is WDE with respect to $\bx$;
    \item[(2)] if $\bx$ is linearly independent ($\mathcal{F}(C)$ is linearly independent for all $C \in \cP$), then $\cP$ is SEL. 
\end{itemize}
\end{theorem}

\begin{proof}
    (1) Let $i, j \in C \in \cP$. Since the solution is $\cP$-synchronised, $\bx_i = \bx_j$ and so $\dot\bx_i = \dot\bx_j$, which implies 
    \begin{align}
        \bh_i(\bx_1,\ldots,\bx_N) = \bh_j(\bx_1,\ldots,\bx_N) \implies
        \bh_i^\text{ext}(\bx_1,\ldots,\bx_N) = \bh_j^\text{ext}(\bx_1,\ldots,\bx_N),
    \end{align}
    since $\bh_i^{C,\ldots, C} = \bh_i^{C,\ldots, C} = 0$ as the coupling functions are non-invasive $\bg^{(m)}(\bx,\ldots,\bx)=0$. 
    \\[5pt]
    (2) Write $\bx_C=\bx_i$ whenever $i\in C$. Let $i, j \in C$, then $\bx_i = \bx_j$ so $\dot\bx_i = \dot\bx_j$ which implies (substitute differential equation, equate, rearrange, write $\bx_l$ as $\bx_{C'}$ whenever $l \in C'$)
    \begin{align}
        \sum_{m=1}^M \sigma_m \sum_{j_1,\ldots,j_m=1}^N \left( A^{(m)}_{i,j_1,\ldots,j_m} - A^{(m)}_{j,j_1,\ldots,j_m}\right) \bg^{(m)}(\bx_C, \bx_{C_1},\ldots,\bx_{C_m}) = 0 \iff \\
        \sum_{m=1}^M \sigma_m \left( h_i^{C_1,\ldots,C_m} - h_j^{C_1,\ldots,C_m} \right) \bg^{(m)}(\bx_C, \bx_{C_1},\ldots,\bx_{C_m}) = 0 \stackrel{\mathcal{F}(C)}{\iff} \\
        \sigma_m \left( h_i^{C_1,\ldots,C_m} - h_j^{C_1,\ldots,C_m} \right) = 0 \stackrel{\sigma_m \neq 0}{\iff}
        h_i^{C_1,\ldots,C_m} = h_j^{C_1,\ldots,C_m}
    \end{align}
    for all $m$, $C_1, \ldots, C_m$, that is, the partition is SEL. 
\end{proof}

As in the Main Text, we define the \emph{quotient} of the hypergraph dynamical system \eqref{eq:dynamical-system-general} and \eqref{eq:coupling-hypergraph-SI} with respect to a partition $\cP$ of the node set $\{1,\ldots,N\}$ into $K = |\cP|$ clusters as the dynamical system with dynamical units $\by_1,\ldots, \by_K$ and $K$ differential equations 
\begin{align}
    \label{eq:quotient-dynamics-hypergraph-1}
    \dot{\by}_i & = \bbf(\by_i) + \bh_i(\by_1,\ldots,\by_K), \quad \text{ where}\\
    \label{eq:quotient-dynamics-hypergraph-2}
    \bh_i(\by_1,\ldots, \by_K) &= \sum_{m=1}^M \sigma_m \sum_{j_1, \ldots, j_m =1}^K B^{(m)}_{i,j_1,\ldots,j_m}\, \bg^{(m)}(\by_i,\by_{j_1},\ldots,\by_{j_m}) \quad \text{ and }\\
    \label{eq:quotient-dynamics-hypergraph-3}
    B^{(m)}_{ij_1\ldots j_m} &= \frac{1}{|C_i|} \sum_{k \in C_i, k_1 \in C_{j_1},\ldots, k_m \in C_{j_m}} A^{(m)}_{k,k_1,\ldots,k_m},
\end{align}
the quotient tensor of $A^{(m)}$ with respect to $\cP$. 

\begin{theorem}\label{thm:quotient-hypergraph}
    Consider the hypergraph dynamical system given by Equations \eqref{eq:dynamical-system-general} and \eqref{eq:coupling-hypergraph-SI}, which we refer to as the parent dynamical system, a partition of the node set $\cP$, and the associated quotient dynamical system given by Equations \eqref{eq:quotient-dynamics-hypergraph-1} to \eqref{eq:quotient-dynamics-hypergraph-3}. Suppose that $\cP$ is SEL. Then,  $\bx=(\bx_1,\ldots,\bx_N)$ is a $\cP$-synchronised solution of the parent dynamical system if and only if $\by=(\by_1,\ldots,\by_K)$ is a solution of the quotient dynamical system, where $\by_k = \bx_i$ whenever $i \in C_k$. 
\end{theorem}

\begin{proof}
    Note that the condition $\by_k = \bx_i$, $i \in C_k$, uniquely defines $\by$ from $\bx$ (since $\bx$ is $\cP$-synchronised), and $\bx$ from $\by$ (since $\cP$ is a partition of the node set).
    Using the equitability condition (SEL), we can write 
    \begin{align}
        B_{kl_1\ldots l_m}^{(m)} 
        &= 
        \frac{1}{|C_k|} \sum_{ i \in C_k, j_1 \in C_{l_1}, \ldots, j_m \in C_{l_m}} A^{(m)}_{ij_1\ldots j_m} = 
        \frac{1}{|C_k|} \sum_{i \in C_k} \overbrace{\sum_{j_1 \in C_{l_1}, \ldots, j_m \in C_{l_m}} A^{(m)}_{ij_1\ldots j_m}}^{=h_i^{C_{l_1},\ldots, C_{l_m}}} \nonumber\\ 
        &\stackrel{\ref{eq:hypergraph-SEL}}{=} 
        \frac{1}{|C_k|} |C_k| h_i^{C_{l_1},\ldots, C_{l_m}} = 
        h_i^{C_{l_1},\ldots, C_{l_m}} =
        \sum_{j_1 \in C_{l_1},\ldots, j_m \in C_{l_m}} A^{(m)}_{ij_1,\ldots,j_m},\label{eq:quotient-last-hypergraph}
    \end{align}
    for any choice $i \in C_k$ and any indices $l_1,\ldots,l_m$ not all equal to $k$.  
    Suppose that $\bx=(\bx_1,\ldots,\bx_N)$ and $\by=(\by_1,\ldots,\by_K)$ satisfy $\by_k = \bx_i$ whenever $i \in C_k$. Let $k \in \{1,\ldots, K\}$ and $i \in C_k$. Then 
    \begin{align}
        \bh_k(\by_1,\ldots,\by_K) 
        &= \sum_{m=1}^M \sigma_m \sum_{l_1, \ldots, l_m =1}^K B^{(m)}_{k,l_1,\ldots,l_m}\, \bg^{(m)}(\by_k,\by_{l_1},\ldots,\by_{l_m})\\ 
        &\stackrel{\eqref{eq:quotient-matrix-equitability}}{=} 
        \sum_{m=1}^M \sigma_m \sum_{l_1, \ldots, l_m =1}^K
        \sum_{j_1 \in C_{l_1},\ldots, j_m \in C_{l_m}} A^{(m)}_{ij_1,\ldots,j_m}
        \, \bg^{(m)}(\by_k,\by_{l_1},\ldots,\by_{l_m})
        \\
        &= 
        \sum_{m=1}^M \sigma_m \sum_{l_1, \ldots, l_m =1}^K
        \sum_{j_1 \in C_{l_1},\ldots, j_m \in C_{l_m}} A^{(m)}_{ij_1,\ldots,j_m}
        \, \bg^{(m)}(\bx_i,\bx_{j_1},\ldots,\bx_{j_m})
        \\
        &= 
        \sum_{m=1}^M \sigma_m 
        \sum_{j_1\ldots, j_m =1}^N A^{(m)}_{ij_1,\ldots,j_m}
        \, \bg^{(m)}(\bx_i,\bx_{j_1},\ldots,\bx_{j_m})
        = \bh_i(\bx_1,\ldots, \bx_N). \label{eq:quotient-last-hypergraph-last}
    \end{align}
    (Note that the second equality follows from \eqref{eq:quotient-matrix-equitability} for $l_1,\ldots,l_m$ not all equal to $k$ and from the non-invasiveness of $\bg^{(m)}$ for $l_1=\ldots=l_m=k$.)
    All in all, if $i \in C_k$, the differential equation for $\bx_i$ in the parent dynamical system is the same as the differential equation for $\by_k$ in the quotient dynamical system. This completes the proof. 
\end{proof}

\subsection{General coupling}\label{sec:general-coupling}
Here, we describe, section by section, the necessary changes to generalise our results so far when we remove the non-invasiveness condition. 

\medskip

\noindent\textbf{Section `Dynamical system and synchronisation'\ref{sec:dynamical_system}} The dynamical systems Eqs.~\eqref{eq:dynamical-system-general}, \eqref{eq:coupling-multilayer-SI} and \eqref{eq:coupling-hypergraph-SI} are the same, removing the assumption $\bg^{(m)}(\bx,\ldots,\bx)= \mathbf{0}$ (last bullet point after Eq.~\eqref{eq:coupling-hypergraph-SI}). 

\medskip

\noindent\textbf{Section `Dynamical and structural equitability'\ref{sec:dynamical-equitability}} In Definition \ref{def:equitability-multiplex}, substitute $\bh_i^{V \setminus C}$ and $h_i^{V \setminus C}$ by $\bh_i$ respectively $h_i$ in Eqs.~\eqref{eq:cluster-dynamically-equitable} and \eqref{eq:cluster-structurally-equitable}, and remove the condition `$C \neq C'$', that is, immediately before Eqs.~\eqref{eq:DE}, \eqref{eq:SE}, \eqref{eq:DEL} and \eqref{eq:SEL}. Ignore Remark \ref{rmk:external}. Theorem \ref{thm:structural-dynamical-equitability} holds verbatim.

\medskip

\noindent\textbf{Section `Linear independence over trajectories'\ref{sec:linear-independence-trajectories}} Remove `$C \neq C'$' from Definition \ref{def:families}, the first bullet point below Eq.~\eqref{eq:families-inclusions}, and the statement of Theorem \ref{thm:equitability-and-linear-independence}. The proof of this theorem is identical, except removing the condition `$C \neq C'$'.

\medskip

\noindent\textbf{Section `Synchronisation and equitability'\ref{sec:synchronisation-and-equitability}} Theorem \ref{thm:cluster-synchronisation-implies-equitability-multilayer} holds but the proof of case (1) now goes as\\[3pt]
\noindent (1) Let $C \in \cP$, $i, j \in C$. Since $\bx=(\bx_1,\ldots,\bx_N)$ is a solution the dynamical system, 
    \begin{equation}
        \dot{\bx}_i = \mathbf{f}(\bx_i) + \bh_i(\bx_1,\ldots,\bx_N) = \mathbf{f}(\bx_i) + \bh_i, 
    \end{equation}
    and similarly for $j$. Since $\bx$ is $\cP$-synchronised, $\bx_i = \bx_j$. In particular, 
    \begin{align}
        \dot{\bx}_i = \dot{\bx}_j \implies \bh_i = \bh_j 
    \end{align}
    and $\cP$ is WDE with respect to $\bx$ by definition.\\[3pt]
Theorem \ref{thm:quotient-multilayer} holds verbatim with the proof adapted by removing the references to $k \neq l$, that is, remove `and any $k \neq l$' and the note in brackets after Eq.~\eqref{eq:quotient-last}. 

\medskip

\noindent\textbf{Section `Synchronisation and equitability for hypergraphs'\ref{sec:dynamics-hypergraphs}} In Definition \ref{def:equitability-hypergraphs}, remove both superscripts `$\textit{ext}$' from Eq.~\eqref{eq:external-hypergraph}, and the definition of $\bh_i^{\text{ext}}$ immediately after, that is, `and we write\ldots'. Ignore Remark \ref{rmk:external-hypergraph}. In Definition \ref{def:SEL-hypergraph}, remove `not all equal to $C$'. Theorem \ref{thm:structural-dynamical-equitability-hypergraph} holds, with verbatim proof. In Definition \ref{def:families-hypergraph}, remove `not all equal to $C$' from the definitions of $\mathcal{F}(C)$ and $\mathcal{F}(C,m)$. In Theorem \ref{thm:equitability-and-linear-independence-hypergraph}, remove `not all equal to $C$' in statement (1). Theorem \ref{thm:cluster-synchronisation-implies-equitability-hypergraph} holds, with the proof of (1) simplified to \\[3pt]
\noindent(1) Let $i, j \in C \in \cP$. Since the solution is $\cP$-synchronised, $\bx_i = \bx_j$ and so $\dot\bx_i = \dot\bx_j$, which implies 
    \begin{align}
        \bh_i(\bx_1,\ldots,\bx_N) = \bh_j(\bx_1,\ldots,\bx_N).
    \end{align}\\[3pt]
Theorem \ref{thm:quotient-hypergraph} holds, with identical proof except deleting `not all equal to $k$' after Eq.~\eqref{eq:quotient-last-hypergraph}, and the note in brackets after Eq.~\eqref{eq:quotient-last-hypergraph-last}.

\subsection{Higher-order explosive synchronisation}
Since equitably (external, or general) is a ‘per layer’ concept, the number of equations that need to be simultaneously satisfied to achieve equability increases (linearly) with the number of layers. It is worse than that: many networks have equitable partitions with high probability somewhere on the network (for example, orbit partitions in real-world networks \cite{macarthur2008symmetry, sanchez2020exploiting}) but for those equitable clusters to occurs on the same nodes in two, or more, layers, it is much less likely. 

For a concrete calculation, suppose that $n_k$ is the average number of clusters with $k$ nodes in the, say for simplicity, coarsest equitable partition for a family of graphs. If we write $p_k$ for the probability of a subset of $k$ nodes to form an equitable cluster, we have the relation $p_k \binom{n}{k} = n_k \le \frac{n}{k}$ hence $p_k \le \frac{n}{k \binom{n}{k}}$ typically very small. On two layers, the probability of the same set of nodes to form an equitable cluster goes from $p_k$ to $p_k^2$ and the expected number of equitable clusters with $k$ nodes from $p_k \binom{n}{k} = n_k$ to $p_k^2 \binom{n}{k}=p_k n_k$. Similarly, for $M$ layers, it goes to $p_k^M \binom{n}{k}=p_k^{M-1} n_k$. That is, $p_k^{M-1}$ is the expected proportion of equitable clusters in one layer that survive after adding $M-1$ layers (sampled from the same family of graphs). Even assuming the largest possible $p_k$ (equal to $n/k$), we have $p_k = \frac{n}{k \binom{n}{k}}$, which decays very quickly to 0 with $n$ and $k$.  

For a concrete example, if $n=100$ and $k = 2, 3, 4$, the values of $p_k$ are $1/99$, $1/4851$, and $1/156849$; for $n=1,000$, these are $1/999$, $1/498501$, and $1/165668499$. Even if the original network has the maximum number of equitable clusters with $k$ nodes (50, 33,  25 respectively 500, 333, 250), the expected number of simultaneously equitable clusters with $k$ nodes on just 2 layers would be 0.5, 0.007, and 0.00002 (1 s.f.) and the same for $n=1,000$ (the expected numbers are the same in the limit $n \to \infty$).

\subsection{Global synchronisation and equitability}
In the non-invasive coupling case, there is always a globally synchronised solution, as explained in the Main Text. This agrees with our structural results: the partition into one cluster is always externally equitable, and a globally synchronised solution is always linearly independent. 

In the general coupling case, we have a similar result, but in this case requiring the (general) equitability of the multiplex network, or hypergraph, and the linear independence of the globally synchronised solution. 

\begin{theorem}
    Consider the multilayer or the hypergraph dynamics, with the same terminology and notation as in Section \ref{sec:dynamical_system}. Then,
    \begin{itemize}
        \item[(1)] if there is a linearly independent, globally synchronised solution of the dynamical system, then the partition into a single cluster is equitable; 
        \item[(2)] if the partition into a single cluster is equitable, and the corresponding quotient dynamics has a solution, then the system has a globally synchronised solution. 
    \end{itemize}
\end{theorem}
\begin{proof}
    (1) A globally synchronised solution is the same as a $\cP$-synchronised solution where $\cP$ is the partition into one cluster. By our results, $\cP$ must be WDE which implies SEL (called equitability in the Main Text in the general coupling case) by the linearly independent hypothesis.\\[2pt]
    (2) Conversely, if the partition is equitable, any solution $\by_s$ of the quotient dynamics (which has the form $\dot\by = \mathbf{f}(\by) + \bh(\by)$, a single equation, on a single quotient node) is a solution of the parent dynamics by repeating the entries of each node, that is, a globally synchronised solution.
\end{proof}
In a network (one layer), the partition into one cluster is equitable is equivalent to a regularity condition, namely that the weighted out-degree on each layer $m$, or for each order $m$, is the same for all nodes. In matrix, or tensor, form, the multiplex, or hypergraph, is \emph{layer-by-layer regular} if for all $i, j \in V$ and $1 \le m \le M$ we have $d_i^{(m)} = d_j^{(m)}$, where
\begin{align}
    d_k^{(m)} &= \sum_{l \in V} A^{(m)}_{kl} = h_k^{V,m} &\text{(multiplex case)}\\
    d_k^{(m)} &= \sum_{l_1,\ldots,l_m \in V} A^{(m)}_{kl_1\ldots l_m} = h_k^{V,\ldots,V} &\text{(hypergraph case)}
\end{align}
for each $k \in V$, the weighted out-degree of node $k$ on layer, respectively for order, $m$.

\subsection{Finding equitable partitions}
Here, we explain how to adapt the Weisfeiler-Lehman (WL) refinement algorithm to other combinatorial structures on a set of nodes, such as multiplex and hypergraphs. We first explain the network (graph) case. 

Consider the graph (network) $G=G(A)$ associated to a given $N \times N$ matrix $A$: it has vertex set $V=\{1,\ldots,N\}$ and a directed edge from $i$ to $j$ ($i, j \in V$) with weight $A_{ji} \neq 0$, and no such edge when $A_{ji} = 0$ (we reverse the indices for convenience). A \emph{connection function} on $G$ is a function $c \colon V \times \mathcal{P}(V) \to X$ where $\mathcal{P}(V)$ is the power set of $V$ (the set of all subsets of $V$) and $X$ is a set. Our main example is the weighted in-degree, that is, the sum of the weights of edges from a subset (cluster) $C$ to a vertex $v$,
\begin{align}
    c(i,C) = \sum_{j \in C} A_{ij}.
\end{align}
A partition $\cP$ is \emph{externally $c$-equitable} if for all $C, C' \in \cP$, $C \neq C'$, and all $i,j \in C$, we have $c(i,C')=c(j,C')$. We define \emph{$c$-equitable} partition similarly, removing the condition $C \neq C'$. 

Given two partitions $\cP$ and $\cP'$ of the same set, we call $\cP'$ \emph{finer than} $\cP$, and $\cP$ \emph{coarser than} $\cP'$, if for all $C' \in \cP'$ there exists $C \in \cP$ such that $C' \subseteq C$. This is an order relation among all partitions of a set. Given a partition $\cP$ and nodes $i,j \in V$, we write $i \sim_\cP j$ if nodes $i$ and $j$ belong to the same cluster in $\cP$.  

The refinement algorithm finds the coarsest $c$-equitable partition of a given partition, and it works as follows. Given a partition $\cP$ and a connection function $c$, we define a refinement (a finer partition) $\cP'$ as the equivalence classes of the equivalence relation $i \sim j$ if $i \sim_{\cP} j$ and $c(i,C) = c(j,C)$ for all $C \in \cP$. If $\cP'=\cP$, then the partition is $c$-equitable; otherwise, we repeat the same procedure on the new partition $\cP'$. The algorithm stops when it finds a $c$-equitable partition or when it reaches the singleton partition, which is always $c$-equitable. The same algorithm applies for external equitability, by redefining the equivalent relation as: $i \sim j$ if $i \sim_{\cP} j$, say $i, j \in C' \in \cP$, and $c(i,C) = c(j,C)$ for all $C \in \cP$, $C \neq C'$, that is, we ignore the internal connectivity on each cluster. 

Now it should be clear that the refinement algorithm can be adapted to multiplex networks, by defining an appropriate connection function. Namely, we define $c(i,C) = (c_1(i,C),\ldots,c_M(i,C))$ where
\begin{align}
     c_m(i,C) = \sum_{j in C} A^{(m)}_{ij} = h_i^{C,m}.
\end{align}
The hypergraph case is a bit more complicated, as we need to define connection functions of the form $c_{m}(i,C_1,\ldots,C_m)$, namely, 
\begin{align}
    c_m(i,C_1,\ldots,C_m) = \sum_{j_1 \in C_1, \ldots, j_m \in C_m} A^{(m)}_{ij_1,\ldots,j_m} = h_i^{C_1,\ldots,C_m}
\end{align}
and substitute the equivalence relation above by $i \sim j$ if $i \sim_\cP j$ and $c_m(i,C_1,\ldots,C_m)=c_m(j,C_1,\ldots,C_m)$ for all $m$ and all $C_1,\ldots, C_m \in \cP$, not all equal to $C$ (for external equitability only). With this changes, the refinement algorithm still works for multiplex networks and hypergraphs. The maximum number of iterations is $N-1$, and in each iteration the column sum of $M$ matrices, respectively tensors, are computed, resulting on a polynomial time algorithm. It is possible to show that the refinement algorithm finds the coarsest $c$-equitable, respectively externally $c$-equitable, partition finer than $\cP$, but this is outside the scope of the present article. 

\subsection{Natural coupling example}
Inspired by the example of natural coupling in \cite{gambuzza2021stability}, consider the hypergraph dynamical system with coupling functions $\bg^{(1)}(x,y) = y^2 - x^2$ and $\bg^{(2)}(x,y,z)=yz-x^2$. Then $\bg^{(2)}(x,y,y)=\bg^{(1)}(x,y)$ so the linear independence condition is never satisfied when $x=x_i$, $y=x_j$, $z=x_k$ and nodes $j$ and $k$ are in the same synchronised cluster, as the coupling functions $\bg^{(1)}$ and $\bg^{(2)}$ would be linearly dependent (in fact, equal). Indeed, for any non-trivial cluster solution, we can take $i \in C$, $j, k \in C'$ and then $\bg^{(2)}(x_C,x_{C'},x_{C'})=\bg^{(1)}(x_C,x_{C'})$, which can never satisfy the linear independence condition. This is an example where our approach would not work and is due to the fact that the coupling functions are not `generic' enough. In particular, in such system, we could have cluster synchronisation without equitability of the underlying hypergraph partition.

\section{Numerical Simulations}

In this section, we comment in more detail on the different simulations displayed in the Main Text, both for the multilayer as well as for the hypergraph case.

\subsection{Multilayer case}
We simulated Lorenz \cite{lorenz1963deterministic} oscillators $\bbf(\bx) = (\sigma(y-z), x(\rho-z)- y, xy-\beta z)$ with parameters $\sigma = 10,\, \rho = 28,\, \beta = 2$ and Laplacian couplings in both layers 1 and 2, in the form $\mathbf{g}^{(1)}(\bx_j,\bx_i) = (x_j - x_i, 0, 0),\; \mathbf{g}^{(2)}(\bx_j,\bx_i) = (0, y_j - y_i, 0)$, where $\bx_i=(x_i,y_i,z_i)$ and $\bx_j=(x_j,y_j,z_j). $

The dynamics were numerically computed in Julia with an RK45 algorithm, up to final time $t_f = 1000$ with a timestep of $\Delta t=10^{-3}$. The synchronization error $S_k$ of the $k$th cluster is computed at late times ($t\in[900,1000]$ at intervals of $0.25$) as the time-averaged root-mean-square deviation defined by
\begin{equation}\label{eq:syncerror-SI}
    S_k = \left\langle \sqrt{\frac{1}{N_k} \sum_{i\in C_k} \left|\bx_i - \overline{\bx}_{k}\right|^2} \right\rangle_{T},
\end{equation}
where $C_k$ is the set of nodes contained in cluster $k$, $\overline{\bx}_{k}$ is the ensemble average of $\bx= (x_i)_{i \in V}$ within $C_k$, and $\langle.\rangle_{T}$ denotes the temporal average over the late time window $T$.

\begin{figure}[ht]
    \centering
    \includegraphics[width=0.8\linewidth]{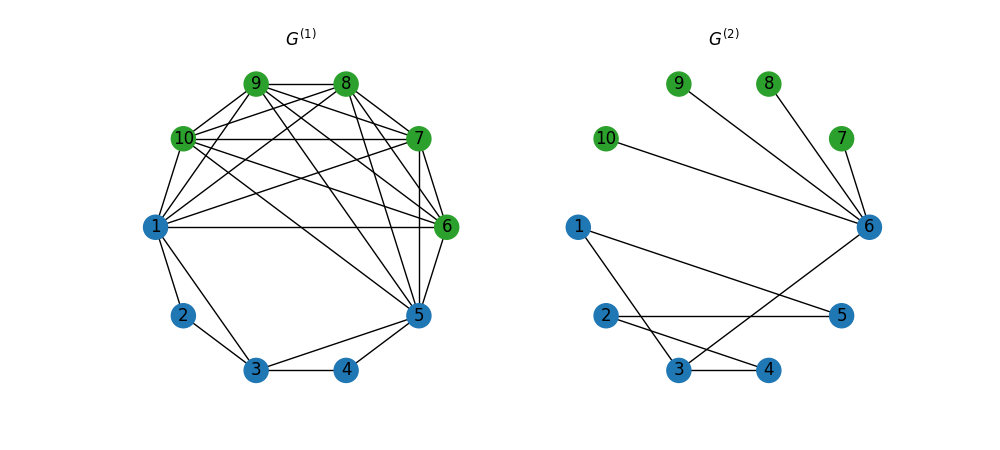}
    \caption{Multilayer network $G_A$ with the same non-trivial external equitable (or structurally equitable) partition in both layers.}
    \label{fig:yescluster_graphs}
\end{figure}

The multiplex network $G_A=\{G_A^{(1)},G_A^{(2)}\}$, for which both layers simultaneously satisfy the structural equitability condition, is shown in Fig.~\ref{fig:yescluster_graphs}: nodes $\{7, 8, 9, 10\}$ form a cluster which is structurally equitable with all other nodes in both layers. As shown in the Main Text, this cluster synchronises for smaller values of coupling strengths $(\sigma_1, \sigma_2)$ than the whole network.

\begin{figure}[ht]
    \centering
    \includegraphics[width=0.8\linewidth]{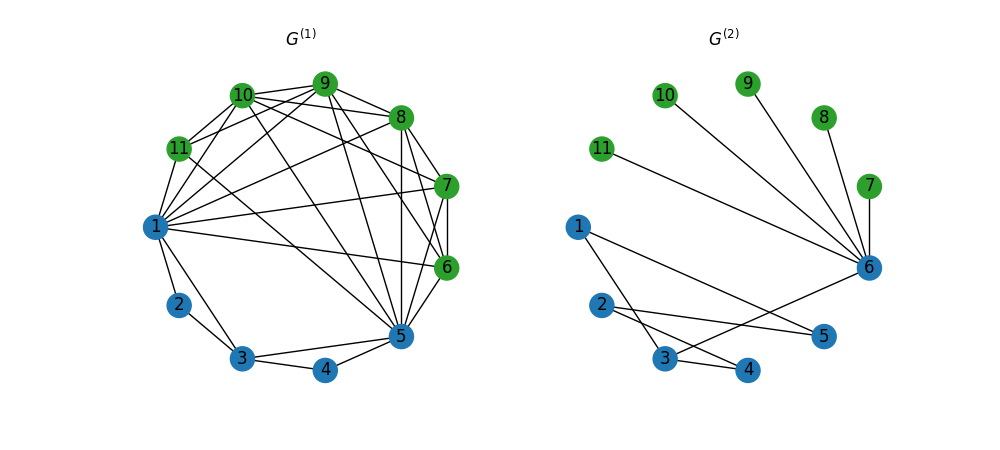}
    \caption{Multilayer network $G_B$ which does not have any common non-trivial external equitable (or structurally equitable) partition in both layers.}
    \label{fig:explosive_example_graphs}
\end{figure}

The multilayer network $G_B=\{G_B^{(1)},G_B^{(2)}\}$,  which does not satisfy the structural equitability condition, is shown in Fig.\ref{fig:explosive_example_graphs}. Although there are non-trivial structurally equitable partitions for each layer separately, with almost identical nodes ($\{6, 7, 8, 9, 10, 11\}$ for the first layer and $\{7, 8, 9, 10, 11\}$ for the second), there are no common structurally equitable partition for both layers. As it was shown in the Main Text, in this case there is no cluster synchronization before total synchronisation when both layers are active.

\subsection{Hypergraph case}

We now consider hypergraph dynamics. We simulated Lorenz \cite{lorenz1963deterministic} oscillators as in the multilayer case, with parameters $\sigma = 10,\, \rho = 28,\, \beta = 8/3$. For simplicity, we assume the couplings to be diffusive, i.e.~$\bg^{(1)}(\bx_i,\bx_j)=\bh^{(1)}(\bx_j) - \bh^{(1)}(\bx_i)$ and $\bg^{(2)}(\bx_i,\bx_j,\bx_k)=\bh^{(2)}(\bx_j,\bx_k) - \bh^{(2)}(\bx_i,\bx_i)$, and so-called \textit{natural coupling} $\bh^{(2)}(\bx,\bx)=\bh^{(1)}(\bx)$. This means that the coupling from pairwise and triadic interactions are essentially the same: a three-body interaction with two nodes on the same state is equivalent to a two-body interaction.

Under these assumptions, we chose the two-body and three-body couplings as $\bh^{(1)}(\bx_j)=(x_j^3,0,0)^T$ and $\bh^{(2)}(\bx_j,\bx_k)=(x_j^2 x_k,0,0)^T$ (cf.~\cite{gambuzza2021stability} where the Master Stability Function for this system is calculated, see Figure 5a in loc.~cit.)

The computation was performed in C with a RK45 algorithm, up to final time $t_f = 1000$ with a timestep of $\Delta t=10^{-3}$. The synchronization error $S_k$ of the $k$th cluster is computed over the last 500 timesteps. The synchronization error is again defined by equation \eqref{eq:syncerror-SI}.

As for the hypergraph itself, it is defined as follows (we give a text description as a visualisation is probably not very insightful). The hypergraph has 20 nodes $V=\{1,...,20\}$ and 4 non-trivial clusters: 
\begin{itemize}
    \item Cluster $C_1 = \{ 1, 2, 3\}$ with connectivity 
    \begin{itemize}
        \item Pairwise: internal edges (1, 2) and (2, 3), external edges ($i$, 15), ($i$, 17), ($i$, 18), for each $i \in C_1$;     
        \item Triadic: no internal hyperedges, external hyperedges (1, 17, 20) and (2, 15, 19).    
    \end{itemize}
    
    (In this cluster, the pairwise structural equitability condition is satisfied, but not the triadic one.)

    \item Cluster $C_2 = \{ 4, 5, 6, 7\}$ with connectivity 
    \begin{itemize}
        \item Pairwise: internal edges (4, 5), (4, 7) and (7, 5), external edges (4, 15) and (7, 20);
        \item Triadic: internal hyperedge (5, 6, 7), external hyperedges ($i$, 18, 20), ($i$, 17, 15) and ($i$, 15, 16) for each $i \in C_2$;
    \end{itemize}

     (In this cluster, the pairwise structural equitability condition is not satisfied, but the triadic one is.)

    \item Cluster $C_3 = \{ 8, 9, 10\}$ with connectivity 
    \begin{itemize}
        \item Pairwise: no internal edges, external edges (8, 15) and (8, 16);
        \item Triadic: internal hyperedge (8, 9, 10), external hyperedges (9, 15, 16) and (10, 15, 16);
    \end{itemize}

    (In this cluster, neither the pairwise nor the triadic structural equitability conditions are satisfied.)

    \item Cluster $C_4 = \{ 11, 12, 13, 14\}$ with connectivity
    \begin{itemize}
        \item Pairwise: internal edges (11, 12), (11, 13) (12, 13) and (13, 14), external edges ($i$, 18) and ($i$, 20) for each $i \in C_4$;
        \item Triadic: internal hyperedge (12, 13, 14), external hyperedges ($i$, 16, 17), ($i$, 19, 20)  for each $i \in C_4$;
    \end{itemize}

    \item The remaining nodes (15 to 20) are not clustered, that is, they are trivial or one-node clusters and there is no additional connectivity among them.
    
\end{itemize}

\end{document}